\documentclass{IEEEtran}
\IEEEtriggeratref{56}

\pdfoutput=1
\usepackage{amsmath}
\usepackage{amsthm}
\usepackage{amssymb}
\usepackage{bm}
\usepackage{xspace}
\usepackage{xcolor}
\usepackage{graphicx}
\usepackage{url}
\usepackage{framed}
\usepackage{float}
\usepackage{rotating}
\usepackage{verbatim}
\usepackage{listings}
\usepackage{lscape}

\usepackage{pgfplots}
\usepackage{pgf}
\usepackage{tikz}
\usetikzlibrary{arrows,shapes.misc,chains,scopes}
\usetikzlibrary{matrix}
\usetikzlibrary{calc}
\usetikzlibrary{fit}
\usetikzlibrary{graphs}
\usepackage{pgfplotstable}
\usepackage{booktabs}
\usepackage{colortbl}

\newcommand{\executeiffilenewer}[3]{%
\ifnum\pdfstrcmp{\pdffilemoddate{#1}}%
{\pdffilemoddate{#2}}>0%
{\immediate\write18{#3}}\fi%
}

\graphicspath{{figures/}}

\theoremstyle{plain}
\newcounter{algocount}
\newcounter{examplecount}

\newcommand{\veca}{\boldsymbol{a}}
\newcommand{\vecb}{\boldsymbol{b}}
\newcommand{\vecB}{\boldsymbol{B}}

\newcommand{\matg}{\boldsymbol{G}}
\newcommand{\matp}{\boldsymbol{P}}
\newcommand{\mati}{\boldsymbol{I}}
\newcommand{\mathh}{\boldsymbol{H}}

\newcommand{\capacity}{\ensuremath{\mathsf{C}}\xspace}

\newcommand{\seta}{\ensuremath{\mathcal{A}}\xspace}

\newcommand{\setx}{\ensuremath{\mathcal{X}}\xspace}

\newcommand{\bmm}{\begin{matrix}}
\newcommand{\emm}{\end{matrix}}
\newcommand{\bpm}{\begin{pmatrix}}
\newcommand{\epm}{\end{pmatrix}}

\newcommand{\bsbm}{\left[\begin{smallmatrix}}
\newcommand{\esbm}{\end{smallmatrix}\right]}
\newcommand{\bspm}{\left(\begin{smallmatrix}}
\newcommand{\espm}{\end{smallmatrix}\right)}

\newcommand{\bbm}{\begin{bmatrix}}
\newcommand{\ebm}{\end{bmatrix}}

\DeclareMathOperator{\expop}{\mathbb{E}}
\DeclareMathOperator{\entop}{\mathbb{H}}
\DeclareMathOperator{\miop}{\mathbb{I}}
\DeclareMathOperator{\kl}{\mathbb{D}}

\newcommand{\oeq}[1]{\overset{\text{(#1)}}{=}}
\newcommand{\ogeq}[1]{\overset{\text{(#1)}}{\geq}}

\usepackage{cite}
\usepackage{printlen}
\usepackage{mdframed}
\usepackage{trsym}
\usepackage{capt-of}
\usepackage{siunitx}
\usepackage{xcolor,colortbl}
\usepackage[normalem]{ulem} % \sout strike out text

\mdfsetup{%
skipabove=1pt,
skipbelow=1pt,
leftmargin=1pt,
rightmargin=1pt,
roundcorner=10pt}

\newtheorem{remark}{Remark}

\newcommand{\bmdrate}{\mathsf{R}_\textnormal{BMD}}
\newcommand{\smdrate}{\mathsf{R}_\textnormal{SMD}}

\title{Bandwidth Efficient and Rate-Matched\\Low-Density Parity-Check Coded Modulation}

\author{Georg~B\"ocherer,~\IEEEmembership{Member,~IEEE,} Patrick~Schulte, and Fabian~Steiner\thanks{G. B\"ocherer and P. Schulte are with the Institute for Communications Engineering, Technische Universit\"at M\"unchen (TUM). F.~Steiner was with the Institute for Communications Engineering, TUM. He is now with the Institute for Circuit Theory and Signal Processing, TUM.

This work was supported by the German Ministry of Education and Research in the framework of an Alexander von Humboldt Professorship. Emails: \texttt{georg.boecherer@tum.de}, \texttt{patrick.schulte@tum.de}, \texttt{fabian.steiner@tum.de}.}
}

\newcommand{\snr}{\ensuremath{\mathsf{SNR}}\xspace}

\DeclareMathOperator{\supp}{supp}

\DeclareMathOperator{\sign}{sign}
\DeclareMathOperator{\bitm}{\pi}

\graphicspath{{figures/}}

\newcommand{\power}{\mathsf{P}}
\newcommand{\heux}{X^\clubsuit}

\newcommand{\heud}{\Delta^\clubsuit}

\newcommand{\nc}{n_{\!\text{c}}}

\newcommand{\kc}{k_{\!\text{c}}}

\newcommand{\dms}[1]{\setlength\fboxsep{1pt}\boxed{#1}}

\pgfplotsset{
width=\columnwidth,
height=0.3\textheight
}

\begin{document}
\maketitle
%%%%%%%%%%%%%%%%%%%%%%%%%%%%%%%
%%%%%%%%%%%%%%%%%%%%%%%%%%%%%%%
%%%%%%%%%%%%%%%%%%%%%%%%%%%%%%%
% ABSTRACT %%%%%%%%%%%%%%%%%%%%
%%%%%%%%%%%%%%%%%%%%%%%%%%%%%%%
%%%%%%%%%%%%%%%%%%%%%%%%%%%%%%%

 \begin{abstract}
A new coded modulation scheme is proposed. At the transmitter, the concatenation of a distribution matcher and a systematic binary encoder performs probabilistic signal shaping and channel coding. At the receiver, the output of a bitwise demapper is fed to a binary decoder. No iterative demapping is performed. Rate adaption is achieved by adjusting the input distribution and the transmission power. The scheme is applied to bipolar amplitude shift keying (ASK) constellations with equidistant signal points and it is directly applicable to two-dimensional quadrature amplitude modulation (QAM). The scheme is implemented by using the DVB-S2 low-density parity-check (LDPC) codes. At a frame error rate of $10^{-3}$, the new scheme operates within less than 1 dB of the AWGN capacity $\frac{1}{2}\log_2(1+\snr)$ at any spectral efficiency between 1 and 5 bits/s/Hz by using only 5 modes,
i.e., 4-ASK with code rate 2/3, 8-ASK with 3/4, 16-ASK and 32-ASK with 5/6 and 64-ASK with 9/10.
\end{abstract}

%\begin{IEEEkeywords}
%\end{IEEEkeywords} 

%%%%%%%%%%%%%%%%%%%%%%%%%%%%%%%
%%%%%%%%%%%%%%%%%%%%%%%%%%%%%%%
%%%%%%%%%%%%%%%%%%%%%%%%%%%%%%%
% INTRODUCTION %%%%%%%%%%%%%%%%
%%%%%%%%%%%%%%%%%%%%%%%%%%%%%%%
%%%%%%%%%%%%%%%%%%%%%%%%%%%%%%%

\section{Introduction}

\begin{figure}
\footnotesize
\centering
\includegraphics{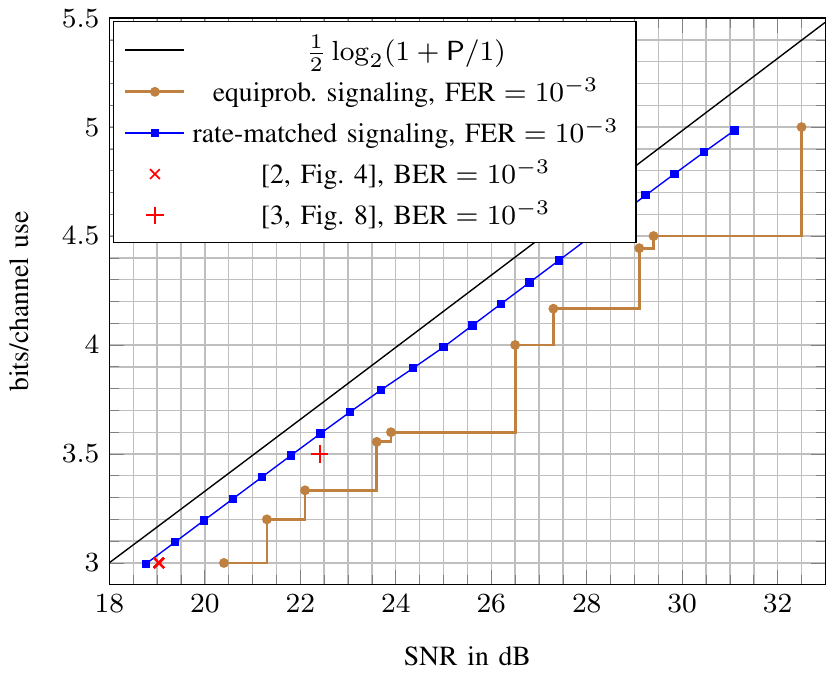}
\caption{Comparison of rate-matching by non-equiprobable signaling to conventional equiprobable signaling. Both schemes use LDPC codes from the DVB-S2 standard. Gains of up to 4 dB (at a spectral efficiency of 4.6 bits/s/Hz) are obtained. The rate-matched scheme uses two modes, namely 64-ASK with a rate 9/10 code and 32-ASK with a rate 5/6 code. Equiprobable signaling uses 12 modes, which combine $\{16,32,64\}$-ASK with $\{3/4,4/5,5/6,8/9,9/10\}$ rate codes. The two schemes are evaluated at a frame error rate (FER) of $10^{-3}$. For comparison, the operating points of the turbo coded modulation schemes with probabilistic shaping suggested in \cite{raphaeli2004constellation} (red $\times$) and \cite{yankov2014rate} (red $+$) are displayed. The operating points of \cite{raphaeli2004constellation,yankov2014rate} have a bit error rate (BER) of $10^{-3}$. Since $\text{BER}\leq\text{FER}$, our rate-matched scheme is more energy efficient than the schemes suggested in \cite{raphaeli2004constellation,yankov2014rate}, see Sec.~\ref{sec:gallager} for a further discussion.}
\label{fig:rateadaption}
\end{figure}

Reliable communication over the additive white Gaussian noise (AWGN) channel is possible if the transmission rate per real dimension does not exceed the capacity-power function
\begin{align}
\capacity(\power)=\frac{1}{2}\log_2(1+\power/1)\label{eq:cpf}
\end{align}
where $\power$ is the transmission power and $\power/1$ is the \emph{signal-to-noise ratio} (SNR). An important consequence of \eqref{eq:cpf} is that to achieve power and bandwidth efficient transmission over a substantial range of SNR, a communication system must adapt its transmission rate to the SNR. Contemporary standards such as DVB-S2 \cite{etsi2009dvb} support input constellations of different sizes and forward error correction (FEC) at various code rates. The combination of a constellation size with a code rate forms a transmission mode. By choosing the appropriate mode, the system adapts to the SNR. In Fig.~\ref{fig:rateadaption} we display the operating points where a frame error rate (FER) less or equal to $10^{-3}$ can be achieved by bipolar \emph{amplitude-shift keying} (ASK) constellations and the DVB-S2 \emph{low-density parity-check} (LDPC) codes. Observe that the SNR gap between the operating points and the capacity-power function varies with the SNR. At the upper corner points, the gap is between 2 and 2.5 dB. Two factors contribute to this gap: first, the LDPC codes have finite length (coding gap) and second, the input distribution is uniform (shaping gap). At the lower corner points, the gap can be as large as 4.5 dB because the system resorts to one of the supported transmission modes.

To provide modes with a finer granularity, we can increase the number of code rates and constellation sizes. This approach is taken in the extension DVB-S2X \cite{etsi2014dvb} of DVB-S2. The system complexity increases with the number of supported code rates, which suggests the use of rate-compatible codes \cite{hagenauer1988rate}. In \cite{li2002rate,ha2004rate,nguyen2012design,chen2015protograph}, rate-compatible LDPC codes are designed.

In this work, we take a different approach and propose a new coded modulation scheme that uses probabilistic shaping to solve the problem of coarse mode granularity and to remove the shaping gap. With ASK constellations and only one LDPC code rate per constellation size,
our scheme operates within less than 1 dB of capacity over the whole considered range of SNR, see Fig.~\ref{fig:rateadaption}. For the transmitter, we propose a new scheme, which we call \emph{probabilistic amplitude shaping} (PAS). At the receiver, bit-metric decoding \cite{ifabregas2008bit,martinez2009bit,bocherer2014achievable} is used. No iterative demapping is required. Our scheme directly applies to two-dimensional \emph{quadrature amplitude modulation} (QAM) constellations by mapping two real ASK symbols to one complex QAM symbol.

This work is organized as follows. In Sec.~\ref{sec:ps}, we give an overview of the related literature on coded modulation with probabilistic shaping. In Sec.~\ref{sec:channel}, we review the information theoretic limits of the AWGN channel and we discuss optimal signaling for ASK constellations. We introduce PAS in Sec.~\ref{sec:ps:encoding} and we show in Sec.~\ref{sec:dm} how PAS can be implemented by distribution matching. In Sec.~\ref{sec:bmd}, we combine PAS at the transmitter with bit-metric decoding at the receiver. We discuss in Sec.~\ref{sec:code} the code design problem posed by our scheme and present a bit-mapper optimization heuristic. We propose a rate adaption scheme in Sec.~\ref{sec:adaption} and we discuss numerical results in Sec.~\ref{sec:numerical}. The work is concluded in Sec.~\ref{sec:conclusions}.

%%%%%%%%%%%%%%%%%%%%%%%%%%%%%%%
%%%%%%%%%%%%%%%%%%%%%%%%%%%%%%%
%%%%%%%%%%%%%%%%%%%%%%%%%%%%%%%
% PROBABILISTIC SHAPING %%%%%%%
%%%%%%%%%%%%%%%%%%%%%%%%%%%%%%%
%%%%%%%%%%%%%%%%%%%%%%%%%%%%%%%

 \section{Related Literature}
\label{sec:ps}
On the additive white Gaussian noise (AWGN) channel, uniformly distributed inputs can be up to 1.53 dB less power efficient than Gaussian inputs \cite[Sec.~IV.B]{forney1984efficient}, see also Sec.~\ref{subsec:shapinggain} of this work. Coded modulation with probabilistic shaping uses a non-uniform distribution on equidistant signal points to overcome the shaping gap. For works with a theoretical focus on the topic, see for example \cite{ling2014achieving,mondelli2014achieve} and references therein. Here, we review some of the existing works that address the practical implementation of probabilistic shaping.

\subsection{Gallager's Scheme}
\label{sec:gallager}
Gallager proposes in \cite[p.~208]{gallager1968information} to use a many-to-one mapping to turn a channel with non-uniform inputs into a super channel with uniform inputs.
This approach is combined with turbo coding in \cite{raphaeli2004constellation} and \cite{yankov2014rate}. In \cite{schreckenbach2005signal}, Gallager's scheme is combined with convolutional codes.
Optimal mappings are investigated in \cite{bocherer2013optimal}, see also \cite[Sec.~I.B]{bocherer2014optimal}.
When Gallager's scheme is combined with binary \emph{forward error correction} (FEC), undoing the many-to-one mapping at the receiver is challenging.
In \cite{raphaeli2004constellation,yankov2014rate,schreckenbach2005signal}, this inversion is achieved by iterative demapping, which increases the system complexity compared to uniform signaling.
In \cite{yankov2014rate}, the authors choose the many-to-one mapping according to the desired transmission rate.
The resulting rate granularity is coarse.
To achieve a finer granularity, more bits must be mapped to the same signal point, which increases the system complexity further.

\subsection{Trellis Shaping}

Trellis shaping is proposed in \cite{forney1992trellis} and it is discussed in detail in \cite[Chap.~4]{tretter2002constellation} and \cite[Sec.~4.4]{fischer2002precoding}. The transmitter first selects a set of sequences. A shaping code then selects from this set the sequence of minimum energy for transmission. The authors of \cite{fischer1998combination} combine multilevel coding with trellis shaping using convolutional codes, see also \cite[Sec~VIII.]{wachsmann1999multilevel}. \emph{Bit-interleaved coded modulation} (BICM) \cite{zehavi1992psk,caire1998bit} is used in \cite{smith2012pragmatic} for lower bit levels and a convolutional shaping code selects the two highest bit-levels for energy minimization. LDPC codes were used as shaping codes in \cite{kaimalettu2007constellation}. Since shaping is achieved by the decoder of a shaping code, it is difficult to make trellis shaping flexible to support different shaping rates.

\subsection{Shell Mapping}

Shell mapping \cite{khandani1993shaping} was proposed independently by several groups around the same time, see \cite[Sec.~VII.]{kschischang1994optimal},\cite[Chap.~8]{tretter2002constellation}. In shell mapping, the transmitter first selects a shell, which is a set of low energy input sequences. It then selects a sequence within the selected shell for transmission. Shell mapping combined with trellis coding is used in the ITU-T Recommendation V.34 \cite{itut1998modem}. The main challenge is to efficiently index the shells.

\subsection{Superposition Coding}

Superposition coding was proposed in \cite{duan1997approaching}. The transmitter uses several encoders. Each encoder maps its output to a signal point. The selected signal points are then superposed and transmitted. The receiver uses multistage decoding. Turbo codes are considered in \cite{duan1997approaching} and variations of superposition coding are proposed and analyzed in \cite{ma2004coded}. The author of  \cite{cronie2010signal} uses LDPC codes and proposes signal constellations that reduce the number of decoding stages. Superposition coding requires sequential decoding of the individual codes or iterative demapping.

\subsection{Concatenated Shaping}

The authors of \cite{legoff2005bit} concatenate BICM encoding with shaping. A turbo encoder is followed by a one-to-one shaping block code. The scheme is improved in \cite{legoff2007constellation} by iterative demapping. A variation of this scheme is presented in \cite{khoo2006bit}, where the one-to-one shaping block code is followed by a many-to-one mapping reminiscent of Gallager's scheme. The shaping scheme \cite{legoff2007constellation} is used in \cite{valenti2012constellation} with LDPC codes on \emph{amplitude and phase-shift keying} (APSK) constellations. Since the shaping decoder is placed before the FEC decoder, only limited gains are observed in \cite{legoff2005bit}. The works \cite{legoff2007constellation,khoo2006bit,valenti2012constellation} therefore use iterative demapping.

\subsection{Bootstrap Scheme}

A bootstrap scheme is proposed in \cite{bocherer2011operating},\cite[Chap.~7]{bocherer2012capacity}, which separates shaping and FEC by first shaping the data and then systematically encoding the shaped data. The generated check bits are also shaped and embedded in the shaped data of the next transmission block. The bootstrap scheme is implemented in \cite{bocherer2011operating},\cite[Chap.~7]{bocherer2012capacity} using Geometric Huffman Coding \cite{bocherer2011matching} for shaping and LDPC codes from the DVB-S2 standard for FEC. Since shaping is done prior to FEC encoding, the bootstrap scheme borrows from the reverse concatenation proposed in \cite{bliss1981circuitry,blaum2007high} for magnetic recording. The authors in \cite{mondelli2014achieve} call the bootstrap scheme \emph{chaining construction} and prove that it is capacity-achieving for any discrete memoryless channel. The drawback of the bootstrap scheme is that shaping is done over several consecutive transmission blocks, which have to be decoded in reverse (bootstrapped). This increases the latency and is prone to error propagation over blocks.

%%%%%%%%%%%%%%%%%%%%%%%%%%%%%%%
%%%%%%%%%%%%%%%%%%%%%%%%%%%%%%%
%%%%%%%%%%%%%%%%%%%%%%%%%%%%%%%
% OPTIMAL SIGNALING %%%%%%%%%%%
%%%%%%%%%%%%%%%%%%%%%%%%%%%%%%%
%%%%%%%%%%%%%%%%%%%%%%%%%%%%%%%

\section{Optimal Signaling over the AWGN Channel}
\label{sec:channel}
\begin{table*}
\footnotesize	
\centering
\caption{SNR gaps in dB of $\heux$ to uniform ASK and capacity $\capacity(\power)$.}
\label{tab:ps:shapinggains}
\begin{tabular}{rrrrrrr}
			&		&$\heux$		&\multicolumn{2}{c}{uniform ASK}	&\multicolumn{2}{c}{capacity $\capacity(\power)$}		\\
constellation		&rate [bits/channel use]	&SNR [dB]		&SNR [dB]		&gap [dB]	&SNR [dB]		&gap [dB]	\\\midrule
4ASK			&1		&4.8180			&5.1181			&-0.3001	&4.7712			&0.0468		\\
8ASK			&2		&11.8425		&12.6187		&-0.7761	&11.7609		&0.0816		\\
16ASK			&3		&18.0910		&19.1681		&-1.0772	&17.9934		&0.0975		\\
32ASK			&4		&24.1706		&25.4140		&-1.2434	&24.0654		&0.1052		\\
64ASK			&5		&30.2078		&31.5384		&-1.3307	&30.0988		&0.1090
\end{tabular}
\end{table*}

The discrete time AWGN channel at time instance $i$ is described by the input-output relation
\begin{align}
Y_i=X_i+Z_i
\end{align}
where the noise terms $Z_i$, $i=1,2,3,\dotsc$ are independent and identically distributed (iid) according to a zero mean Gaussian distribution with variance one. For $\nc$ channel uses, the input is subject to the power constraint
\begin{align}
\frac{\expop\left[\sum_{i=1}^{\nc} |X_i|^2\right]}{\nc}\leq\power
\end{align}
where $\expop[\cdot]$ denotes expectation.
\subsection{AWGN Capacity}
Let $\hat{X}^{\nc}=\hat{X}_1,\hat{X}_2,\dotsc,\hat{X}_{\nc}$ be the output of the decoder that estimates the input $X^{\nc}$ from the output $Y^{\nc}$.
The block error probability is $P_e=\Pr(\hat{X}^{\nc}\neq X^{\nc})$.
The channel coding theorem \cite[Theorem~7.4.2]{gallager1968information} states that by choosing $\nc$ large enough, $P_e$ can be made as small as desired if the transmission rate $R$ is smaller than the channel capacity $\capacity(\power)$.
The goal of this work is to design a modulation scheme that achieves reliable transmission close to the capacity-power function \eqref{eq:cpf}, i.e., for \textbf{any} average signal power $\power$, we want to reliably transmit data over the AWGN channel at a rate that is close to $\capacity(\power)$.

\subsection{Amplitude Shift Keying}
Let $X$ be distributed on some finite alphabet $\setx$ of signal points and suppose $\expop[|X|^2]=\power$. The channel coding theorem states that reliable transmission at rate $R$ with average power $\power$ is possible if
\begin{align}
R<\miop(X;Y)
\end{align}
where $\miop(\cdot;\cdot)$ denotes the mutual information in bits. The first step in designing a coded modulation system is thus to choose an alphabet $\setx$ and a distribution $P_X$ on $\setx$ such that $\miop(X;Y)$ is close to the maximum value $\capacity(\power)$. As input alphabet, we use an ASK constellation with $2^m$ signal points, namely
\begin{align}
\setx=\{\pm 1,\pm 3,\dotsc,\pm (2^m-1)\}.\label{eq:signaling:setx}
\end{align}
We scale $X$ by the \emph{constellation scaling} $\Delta>0$ and the resulting input/output relation is
\begin{align}
Y=\Delta X+Z.\label{eq:ps:model}
\end{align}
The power constraint for $X$ is now
\begin{align}
\expop[|\Delta X|^2]\leq\power\label{eq:ps:askpower}.
\end{align}

\subsection{Optimization of the ASK Input}
\label{subsec:optimizedinput}
We use signal point $x\in\setx$ with probability
\begin{align}
\begin{split}
P_{X_\nu}(x)=A_\nu e^{-\nu |x|^2},\qquad
A_\nu=\frac{1}{\sum_{x'\in\setx} e^{-\nu |x'|^2}}.
\end{split}\label{eq:ps:em}
\end{align}
The scalar $A_\nu$ ensures that the probabilities assigned by $P_{X_\nu}$ add to $1$. Distributions of the form \eqref{eq:ps:em} are called \emph{Maxwell-Boltzmann distributions}, see, e.g., \cite{kschischang1993optimal}. For a fixed constellation scaling $\Delta$ and power constraint $\power$, we choose the input distribution 
\begin{align}
P_{X_\Delta}(x)=&P_{X_\nu}(x)\text{ with }\nu\colon \expop[|\Delta X_\nu|^2]=\power.\label{eq:ps:embisection}
\end{align}
The distribution $P_{X_\Delta}$ maximizes entropy subject to the power constraint \eqref{eq:ps:askpower}, see \cite[Chapter~12]{cover2006elements}. Some basic manipulations show that $\expop[|X_\nu|^2]$ is strictly monotonically decreasing in $\nu$. Thus, the value of $\nu$ for which the condition \eqref{eq:ps:embisection} is fulfilled can be found efficiently by using the \emph{bisection method}\index{bisection method}. For each constellation scaling $\Delta$, the distribution $P_{X_\Delta}$ satisfies the power constraint. We now maximize the mutual information over all input distributions from this family, i.e., we solve
\begin{align}
\max_\Delta \miop(X_\Delta;\Delta X_\Delta+Z).\label{eq:ps:spacingoptimization}
\end{align}
The mutual information $\miop(X_\Delta;\Delta X_\Delta+Z)$ is a unimodal function of $\Delta$ and the optimization problem can be solved efficiently by using the \emph{golden section method}. We denote the maximizing scaling by $\heud$ and the corresponding distribution by $P_{\heux}$.
\subsection{Shaping Gap}
\label{subsec:shapinggain}
In Table~\ref{tab:ps:shapinggains}, we display the shaping gains of our input $\heux$ as compared to a uniformly distributed input. For increasing rates and constellation sizes, the shaping gain increases and approaches the upper bound of $10\log_{10}\frac{\pi e}{6}\approx 1.53$ dB \cite[Sec.~IV.B]{forney1984efficient}. The bound can also be derived by probabilistic arguments, see \cite[Comment 4)]{ozarow1990capacity}. For 64-ASK and a rate of 5 bits per channel use, the shaping gain is $\SI{1.33}{\decibel}$.
\begin{remark}
The distribution $P_{\heux}$ is suboptimal in general.
The optimal distribution $P_{X^*}$ can be approximated numerically by using the Blahut-Arimoto algorithm \cite{blahut1972computation,arimoto1972algorithm}. However, there is no analytical expression of $P_{X^*}$. For the operating points in Table~\ref{tab:ps:shapinggains}, the energy efficiency of $\heux$ is within 0.1 dB of capacity $\capacity(\power)$, so the gain from optimizing further is bounded by 0.1 dB. 
\end{remark}

\subsection{Input Distribution: Operational Meaning}

Suppose we use the channel $\nc$ times, i.e., we transmit length $\nc$ codewords. We use codeword $x^{\nc}$ with probability $P_{X^{\nc}}(x^{\nc})$. If our transmission rate $R$ is larger than  the average mutual information between input and output, i.e., if
\begin{align}
R>\frac{\sum_{i=1}^{\nc}\miop(X_i;Y_i)}{\nc}\label{eq:converse}
\end{align}
then the probability of error $\Pr(\hat{X}^n\neq X^n)$ is bounded away from zero for any decoder. This result follows as a corollary of the converse of the channel coding theorem \cite[Theorem~7.3.1]{gallager1968information}.
% For completeness, we provide a proof in Appendix~\ref{app:converse}.

The mutual information terms $\miop(X_i;Y_i)$ are calculated according to the marginal distributions $P_{X_i}$. We call the condition \eqref{eq:converse} the \emph{Coded Modulation Converse}. The insight we take from \eqref{eq:converse} is that building a transmitter with marginal input distributions
\begin{align}
P_{X_i}\approx P_{\heux}\label{eq:ps:condition}
\end{align}
is necessary to achieve reliable transmission close to the rates displayed for $\heux$ in Table~\ref{tab:ps:shapinggains}.

%%%%%%%%%%%%%%%%%%%%%%%%%%%%%%%
%%%%%%%%%%%%%%%%%%%%%%%%%%%%%%%
%%%%%%%%%%%%%%%%%%%%%%%%%%%%%%%
% PROBABILISTIC AMPLITUDE SHAPING
%%%%%%%%%%%%%%%%%%%%%%%%%%%%%%%
%%%%%%%%%%%%%%%%%%%%%%%%%%%%%%%

\section{Probabilistic Amplitude Shaping}
\label{sec:ps:encoding}

We now develop a transmitter with property \eqref{eq:ps:condition}.

\subsection{Preliminaries}
\label{subsec:pas:preliminaries}
\label{subsec:ucba}
We make the following two observations:
\subsubsection{Amplitude-Sign Factorization} We can write $\heux$ as
\begin{align}
\heux=A\cdot S
\end{align}
where $A=|\heux|$ is the \emph{amplitude} of the input and where $S=\sign (\heux)$ is the \emph{sign} of the input. By \eqref{eq:signaling:setx}, the amplitudes take values in
\begin{align}
\seta:=\{1,3,\dotsc,2^m-1\}.
\end{align}
We see from \eqref{eq:ps:em} that the distribution $P_{\heux}$ is symmetric around zero, i.e., we have
\begin{align}
P_{\heux}(x)=P_{\heux}(-x)
\end{align}
and therefore, $A$ and $S$ are stochastically independent and $S$ is uniformly distributed, i.e., we have
\begin{align}
&P_{\heux}(x)=P_{A}(|x|)\cdot P_S(\sign(x)),\quad \forall x\in\setx\label{eq:pas:pa}\\
&P_S(1)=P_S(-1)=\frac{1}{2}.
\end{align}
\subsubsection{Uniform Check Bit Assumption}
\begin{figure}
\footnotesize
\centering
\includegraphics{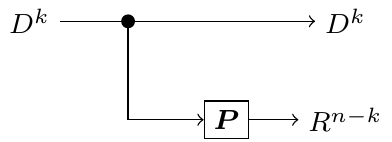}
\caption{Uniform check bit assumption \cite[Sec.~7.1.3]{bocherer2012capacity}: Data bits $D^k=D_1D_2\dotsb D_k$ with some arbitrary distribution $P_{D^k}$ (the data bits are possibly dependent and non-uniformly distributed) are encoded by a systematic generator matrix $[\mati_k|\matp]$ of an $(n,k)$ binary code. The encoder copies the data bits $D^k$ to the output, which preserves the probability distribution of $D^k$. The encoder calculates $n-k$ redundancy bits $R^{n-k}$ by multiplying $D^k$ with the parity matrix $\matp$. Since each $R_i$ is a modulo two sum of several data bits, we assume that the redundancy bits $R^{n-k}$ are approximately uniformly distributed.}
\label{fig:ucba}
\end{figure}
The second observation is on systematic binary encoding. A systematic generator matrix of an $(n,k)$ binary code has the form
\begin{align}
\matg=[\mati_k|\matp]
\end{align}
where $\mati_k$ is the $k\times k$ identity matrix and $\matp$ is a $k\times(n-k)$ matrix. $\matp$ is the \emph{parity matrix} of the code. The generator matrix $\matg$ maps $k$ data bits $D^k$ to a length $n$ codeword via
\begin{align}
D^k\matg=(D^k R^{n-k})
\end{align}
where $R^{n-k}$ are redundant bits that are modulo-two sums of data bits. See Fig.~\ref{fig:ucba} for an illustration. The distribution of the modulo two sum of a large enough number of bits is close to uniform, by a central limit theorem kind of argument. The \emph{uniform check bit assumption} \cite[Sec.~7.1.3]{bocherer2012capacity} states that the marginal distributions $P_{R_i}$ are uniform for a large class of data distributions $P_{D^k}$.

\subsection{Encoding Procedure}
\label{subsec:encoding}
\begin{figure*}
\centering
\footnotesize
\includegraphics{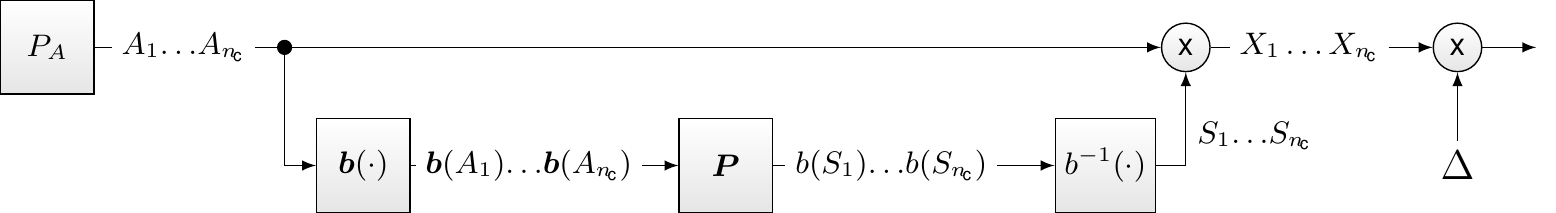}
\caption{PAS. The ASK amplitudes $A_i$ take values in $\seta=\{1,3,\dotsc,2^m-1\}$. The amplitudes $A_i$ are represented by their binary labels $\vecb(A_i)$. Redundancy bits $b(S_i)$ result from multiplying the binary string $b(A_1)b(A_2)\dotsb b(A_{\nc})$ by the parity matrix $\matp$ of a systematic generator matrix $[\mati|\matp]$. The redundancy bits $b(S_i)$ are transformed into signs $S_i$ and multiplied with the amplitudes $A_i$. The resulting signal points $X_i=A_i S_i$ take values in $\setx=\{\pm 1,\pm 3,\dotsc,\pm (2^m-1)\}$. The signal points $X_i$ are scaled by $\Delta$ and $\Delta X_i$ is transmitted over the channel.}
\label{fig:encoder}
\end{figure*}
Consider block transmission with $\nc$ symbols from a $2^m$-ASK constellation. Since we use binary error correcting codes, we label each of the $2^{m-1}$ amplitudes by a binary string of length $m-1$ and we label each of the signs $\pm 1$ by a bit, i.e., we use
\begin{align}
A&\mapsto\vecb(A)\in\{0,1\}^{m-1}\\
S&\mapsto b(S)\in\{0,1\}.
\end{align}
For the sign, we use $b(-1)=0$ and $b(1)=1$. We discuss the choice of $\vecb(A)$ in Sec.~\ref{subsec:labeling}. We use a rate $k/n=(m-1)/m$ binary code with systematic generator matrix $\matg=[\mati_k|\matp]$. For block transmission with $\nc$ channel uses, the block length of the code is $n=\nc m$ and the dimension of the code is $k=\nc(m-1)$. The encoding procedure is displayed in Fig.~\ref{fig:encoder}. It works as follows.
\begin{mdframed}
\begin{enumerate}
\item A \emph{discrete memoryless source} (DMS) $\dms{P_A}$ outputs amplitudes $A_1,A_2,\dotsc,A_{\nc}$ that are iid according to $P_A$. We explain in Sec.~\ref{sec:dm} how the DMS $\dms{P_A}$ can be emulated from binary data by distribution matching. 
\item Each amplitude $A_i$ is represented by its label $\vecb(A_i)$.
\item The resulting length $(m-1)\nc=k$ binary string is multiplied by the parity matrix $\matp$ to generate $n-k=\nc$ sign labels $b(S_1),b(S_2),\dotsc,b(S_{\nc})$.
\item Each sign label $b(S_i)$ is transformed into the corresponding sign $S_i$.
\item The signal $X_i=A_i\cdot S_i$ is scaled by $\Delta$ and transmitted.
\end{enumerate}
\end{mdframed}
We call this procedure \emph{probabilistic amplitude shaping} (PAS). Since the signs $S^{\nc}$ are a deterministic function of the amplitudes $A^{\nc}$, the input symbols $X_1,X_2,\dotsc,X_{\nc}$ are correlated. Let's check if the marginal distributions $P_{X_i}$ fulfill condition \eqref{eq:ps:condition}. We have
\begin{align}
P_{X_i}(x_i)&=P_{A_iS_i}(|x_i|,\sign(x_i))\\
&\approx P_{A_i}(|x_i|)P_{S_i}(\sign(x_i))\label{eq:independence}\\
&=P_{A}(|x_i|)P_{S_i}(\sign(x_i))\\
&\approx P_{A}(|x_i|)\frac{1}{2}\label{eq:uniformity}\\
&=P_{\heux}(x)
\end{align}
where we used the uniform check bit assumption in \eqref{eq:uniformity}. We discuss when we have equality in \eqref{eq:independence} and \eqref{eq:uniformity}.
\begin{itemize}
\item We have equality in \eqref{eq:independence} if $A_i$ and $S_i$ are independent. This is the case if the redundant bit $b(S_i)$ does not depend on the binary label of $A_i$ (the bit $b(S_i)$ can still depend on other $A_j$, $j\neq i$). On the other hand, if $b(S_i)$ is determined by the binary index of $A_i$ only, then $A_i$ and $S_i$ are dependent and we will not have equality in \eqref{eq:independence}.
\item Recall that $b(S_i)$ is the modulo two sum of binary random variables. If one of these binary random variables is independent of the others and uniformly distributed, then a basic probability calculation shows that $b(S_i)$ is exactly uniformly distributed and we have equality in \eqref{eq:uniformity}.
\end{itemize}
For the rest of this work, we assume that our scheme achieves equality in \eqref{eq:independence} and \eqref{eq:uniformity}. Our numerical results support this assumption. Of course, if the empirical results deviate from what theory predicts, we should check the assumptions of equality in \eqref{eq:independence} and \eqref{eq:uniformity}.

\begin{remark}
PAS is a special case of the bootstrap scheme \cite{bocherer2011operating,mondelli2014achieve}. After encoding, the redundancy bits are uniformly distributed. However, instead of transforming them into a sequence of symbols with a non-uniform distribution and transmitting them in the next block as is done in the bootstrap scheme, they can be used directly as sign labels in the same block, since the uniform distribution is already optimal.
\end{remark}
\begin{remark}
PAS can be seen as a probabilistic version of shell mapping. By the weak law of large numbers, 
\begin{align}
\frac{\sum_{i=1}^{\nc}A_i^2}{\nc}\approx \power
\end{align}
with high probability, i.e., the source $P_A$ selects an amplitude sequence in a shell in the $\nc$-dimensional space that contains the sequences of energy $\approx\nc\power$. Multiplying the amplitudes by signs does not change the power, i.e., the transmitted signal $X^{\nc}$ remains in the shell selected by the amplitude source $P_A$.
\end{remark}

\subsection{Optimal Operating Points}
\label{subsec:pas:ops}
\begin{figure}
\footnotesize
%\begin{minipage}[t]{0.48\textwidth}
\centering
\includegraphics{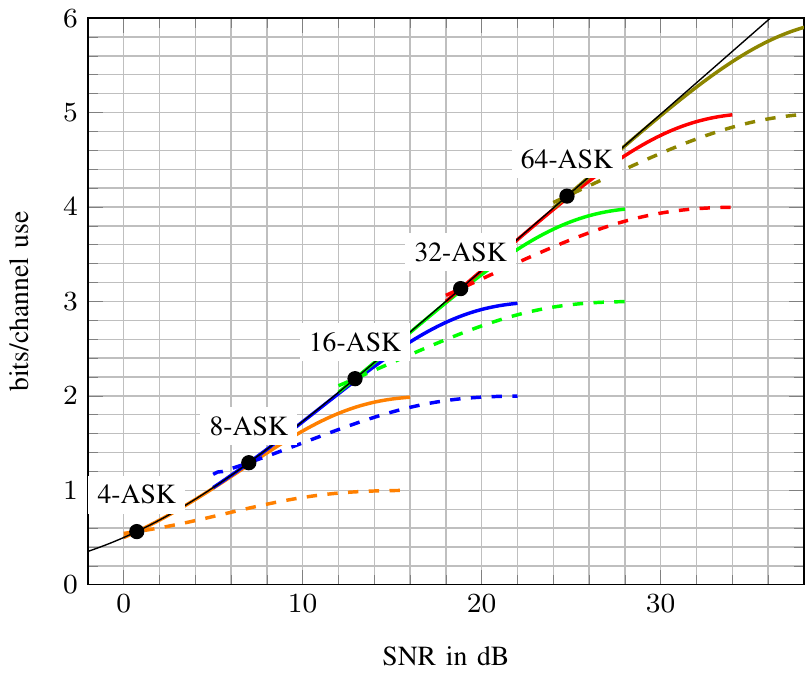}
\caption{The mutual information curves (solid) and the transmission rate curves (dashed) for ASK. The optimal operating points for rate $(m-1)/m$ codes are indicated by dots.}
\label{fig:ops}
%\end{minipage}
%\hfill
%\\[0.5cm]
%\begin{minipage}[t]{0.48\textwidth}
\end{figure}
We study the rates at which reliable transmission is possible with our scheme. By \eqref{eq:converse}, reliable communication at rate $R$ is achievable only if
\begin{align}
R<\frac{\sum_{i=1}^{\nc} \miop(X_i;Y_i)}{\nc}=\miop(X;Y)=\miop(AS;Y).\label{eq:encoderconverse}
\end{align}
Since $A^{\nc}$ represents our data, our transmission rate is
\begin{align}
R=\frac{\entop(A^{\nc})}{\nc}=\entop(A)\quad\left[\frac{\text{bits}}{\text{channel use}}\right]
\end{align}
and condition \eqref{eq:encoderconverse} becomes
\begin{align}
\entop(A)<\miop(AS;Y).
\end{align}
In Fig.~\ref{fig:ops}, both the mutual information $\miop(AS;Y)$ (solid lines) and transmission rate $\entop(A)$ (dashed lines) are displayed for $4,8,16,32$, and $64$-ASK. For high enough SNR, the mutual information saturates at $m$ bits and the transmission rate saturates at $m-1$ bits. Optimal error correction for block length $\nc\to\infty$ would operate where the transmission rate curve crosses the mutual information curve. These crossing points are indicated by dots in Fig.~\ref{fig:ops}. How to operate at other transmission rates is the topic of rate adaption and we discuss this in detail in Sec.~\ref{sec:adaption}.

\subsection{PAS for Higher Code Rates}
\label{sec:ps:op}
\begin{figure}
\centering
\includegraphics{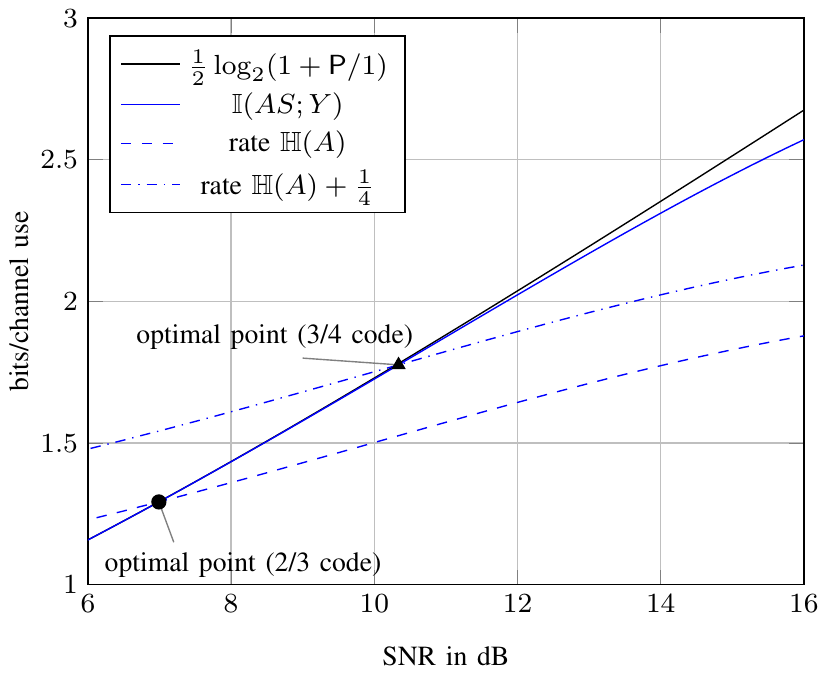}
\caption{Optimal operating points of $8$-ASK for PAS ($c=2/3$) and extended PAS ($c=3/4$).}
\label{fig:ops_extended}
%\end{minipage}
\end{figure}
\begin{figure*}
\centering
\footnotesize
\includegraphics{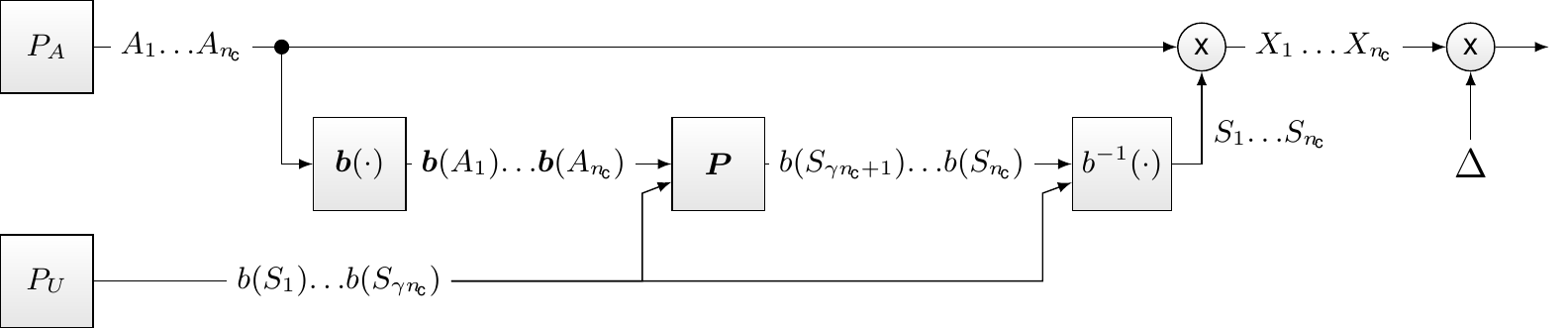}
\caption{Extension of PAS to code rates higher than $(m-1)/m$. The fraction $\gamma$ of the signs is used for data, which is modelled as the output of a Bernoulli-1/2 DMS $P_U$.}
\label{fig:encoderAdapted}
\end{figure*}
%\begin{figure}
%\footnotesize
%\centering
%\input{figures/ops_extended}
%\caption{Optimal operating points of $8$-ASK for PAS ($c=2/3$) and extended PAS ($c=3/4$).}
%\label{fig:ops_extended}
%\end{figure}
We observe in Fig.~\ref{fig:ops} that the ASK mutual information curves stay close to the capacity $\capacity(\power)$ over a certain range of rates above the optimal operating points. We therefore extend our PAS scheme to enable the use of code rates higher than $(m-1)/m$ on $2^m$-ASK constellations. We achieve this by using some of the signs $S_i$ for uniformly distributed data bits. We illustrate this extension of the PAS scheme in Fig.~\ref{fig:encoderAdapted}. Let $\gamma$ denote the fraction of signs used for data bits. We interpret $\gamma\nc$ uniformly distributed data bits as sign labels $b(S_1)\dotsb b(S_{\gamma\nc})$. These $\gamma\nc$ bits and the $(m-1)\nc$ bits from the amplitude labels are encoded by the parity matrix of a rate $c$ code, which generates the remaining $(1-\gamma)\nc$ sign labels. The code rate can be expressed in terms of $m$ and $\gamma$ as  
\begin{align}
c=\frac{m-1+\gamma}{m}.
\end{align}
For a given code rate $c$, the fraction $\gamma$ is given by
\begin{align}
\gamma = 1-(1-c)m\label{eq:gammaofc}.
\end{align}
Since a fraction $\gamma$ of the signs now carries information, the transmission rate of the extended PAS scheme is given by
\begin{align}
R=\frac{\entop(A^{\nc})+\entop(S^{\gamma\nc})}{\nc}=\entop(A)+\gamma\;\left[\frac{\text{bits}}{\text{channel use}}\right].
\end{align}
The optimal operating point is then given by the crossing of the rate curve $\entop(A)+\gamma$ and the mutual information. In Fig.~\ref{fig:ops_extended}, we display for $8$-ASK the optimal operating points for $c=2/3$ and $c=3/4$.

%%%%%%%%%%%%%%%%%%%%%%%%%%%%%%%
%%%%%%%%%%%%%%%%%%%%%%%%%%%%%%%
%%%%%%%%%%%%%%%%%%%%%%%%%%%%%%%
% DISTRIBUTION MATCHING %%%%%%%
%%%%%%%%%%%%%%%%%%%%%%%%%%%%%%%
%%%%%%%%%%%%%%%%%%%%%%%%%%%%%%%

\section{Distribution Matching}
\label{sec:dm}
\begin{figure}
\centering
\footnotesize
\includegraphics{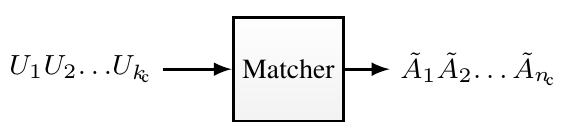}
\caption{The matcher transforms uniform data blocks of length $\kc$ into $\nc$ amplitudes 
that are approximately distributed according to the desired distribution $P_A$. By replacing the amplitude source $P_A$ in the PAS diagrams in Fig.~\ref{fig:encoder} and Fig.~\ref{fig:encoderAdapted} by a matcher, our scheme provides a binary interface to the source coding part of a digital communication system.}
\label{fig:emulator}
\end{figure}
In a digital communication system, usually a binary interface separates the source coding part from the channel coding part \cite[Chap. 1]{gallager2008principles}. Up to now, our scheme does not have such a binary interface, since some of our data is output by an amplitude source $\dms{P_A}$. We therefore add a device that takes as input $\kc$ uniformly distributed independent bits $U^{\kc}$ and outputs $\nc$ amplitudes $\tilde{A}^{\nc}$. We illustrate this in Fig.~\ref{fig:emulator}. We require the following properties for our device:
\begin{itemize}
\item[P1] The input provides a binary interface to the source coding part of the digital communication system.
\item[P2] The output is approximately the output of a DMS $\dms{P_A}$.
\item[P3] The input can be recovered from the output, i.e., the mapping is invertible.
\end{itemize}
A device with such properties is called a \emph{distribution matcher} \cite{bocherer2011matching}. Variable length distribution matchers were proposed in \cite[Sec.~IV.A]{forney1984efficient}, \cite[Sec.~VII]{kschischang1993optimal} and \cite{ungerboeck2002huffman} and their design was studied in \cite{kschischang1993optimal,cai2007probabilistic,bocherer2011matching,bocherer2012capacity,amjad2013fixed,baur2015arithmetic}. 

Variable length matchers can lead to buffer overflow, synchronization loss and error propagation, see, e.g., \cite[Sec.~1]{kschischang1993optimal}. We therefore use the fixed length distribution matcher proposed in \cite{schulte2015constant}. This matcher, called \emph{constant composition distribution matcher} (CCDM), has the properties P1 and P3. To address property P2, we need to say what we mean by ``matcher output $\tilde{A}^{\nc}$ is approximately the output of a DMS $\dms{P_A}$''. By \cite[Theorem~1.2]{wyner1975common}, a good measure for similarity in the context of channel coding is the normalized informational divergence (also known as \emph{Kullback-Leibler divergence} or \emph{relative entropy} \cite[Sec.~2.3]{cover2006elements}) of the output distribution $P_{\tilde{A}^{\nc}}$ and the memoryless distribution $P_{A}^{\nc}$. The normalized informational divergence is
\begin{align}
\frac{\kl(P_{\tilde{A}^{\nc}}\Vert P_A^{\nc})}{\nc}=\frac{\displaystyle\sum_{a^{\nc}\in\supp P_{\tilde{A}^{\nc}}}P_{\tilde{A}^{\nc}}(a^{\nc})\log_2\frac{P_{\tilde{A}^{\nc}}(a^{\nc})}{P_A^{\nc}(a^{\nc})}}{\nc}\label{eq:id}
\end{align}
where $\supp P_{\tilde{A}^{\nc}}$ is the support of $P_{\tilde{A}^{\nc}}$. The CCDM has property P2 in the following sense:
\begin{enumerate}
\item As $\nc \rightarrow \infty$, the normalized informational divergence \eqref{eq:id} approaches zero.
\item As $\nc \rightarrow \infty$, the rate approaches $\entop(A)$, i.e., we have
\begin{align}
\frac{\kc}{\nc}\to\entop(A).
\end{align}
\end{enumerate}
We use the CCDM to emulate the amplitude source $\dms{P_A}$. The output length $\nc$ is finite, and we use the actual rate $\kc/\nc$ for performance evaluations.

%%%%%%%%%%%%%%%%%%%%%%%%%%%%%%%
%%%%%%%%%%%%%%%%%%%%%%%%%%%%%%%
%%%%%%%%%%%%%%%%%%%%%%%%%%%%%%%
% BIT-METRIC DECODING %%%%%%%%%
%%%%%%%%%%%%%%%%%%%%%%%%%%%%%%%
%%%%%%%%%%%%%%%%%%%%%%%%%%%%%%%

 \section{Bit-Metric Decoding}
\label{sec:bmd}

The receiver estimates the transmitted codeword $X^{\nc}$ from the channel outputs $Y^{\nc}$. In this section, we show how this can be implemented by a bit-metric decoder.
\subsection{Preliminaries: Binary Labeling and Decoding}
Consider a coded modulation system where the input $X$ takes values in a $2^m$-ASK constellation. An optimal decoder uses the symbol-metric $p_{Y|X}$ \cite[Sec.~II.B]{martinez2009bit} and can achieve the rate
\begin{align}
\smdrate=\miop(X;Y)\label{eq:cm}
\end{align}
where SMD stands for \emph{symbol-metric decoding} (SMD). We are interested in successfully decoding at a transmission rate close to $\smdrate$ by using a \emph{binary} decoder. Recall that our PAS scheme labels the amplitude $A=|X|$ by a length $m-1$ binary string $\vecb(A)$ and the sign $S=\sign(X)$ by one bit $b(S)$. The length $m$ binary string 
\begin{align}
\vecB=B_1B_2\dotsb B_m:=b(S)\vecb(A)
\end{align}
thus assigns to each signal point $x\in\setx$ a label via
\begin{align}
\text{label}(x)=\text{label}(\sign(x))\text{label}(|x|)=b_1b_2\dotsb b_m.
\end{align}
Since the labeling is one-to-one, we can also select a signal point for transmission by choosing the label $\vecB$, i.e., we use
\begin{align}
X=x_{\vecB}=\{x\in\setx\colon \text{label}(x)=\vecB\}.
\end{align}
We can interpret $\vecB$ as the channel input and the input/output relation of our channel as
\begin{align}
Y=\Delta x_{\vecB}+Z.\label{eq:labelchannel}
\end{align}

Using the chain rule, we expand the mutual information of $\vecB$ and $Y$ as
\begin{align}
\miop(\vecB;Y)&=\sum_{i=1}^m \miop(B_i;Y|B^{i-1})\\
&=\miop(B_1;Y)+\miop(B_2;Y|B_1)+\dotsb
\nonumber\\&\qquad\qquad
+\miop(B_m;Y|B_1\dotsb B_{m-1}).
\end{align}
This expansion suggests the following binary decoding:
\begin{enumerate}
\item Use the channel output $Y$ to calculate an estimate $\hat{B_1}$.
\item Successively use the output $Y$ and the estimates $\hat{B_1}\dotsb\hat{B}_{i-1}$ to calculate the next estimate $\hat{B}_i$. 
\end{enumerate}
This approach is called \emph{multistage decoding} (MD). It requires \emph{multilevel coding} (MLC) at the transmitter, i.e., on each bit-level, an individual binary code with block length $\nc$ is used. MLC/MD was first introduced in \cite{imai1977new} and it is discussed in detail, e.g., in \cite{wachsmann1999multilevel}. To use MLC/MD, we would need to modify our PAS scheme. A simpler approach is to ignore the estimates $\hat{B}_j$, $j\neq i$ when estimating $B_i$. This reduces the mutual information, which can be seen as follows
\begin{align}
\miop(\vecB;Y)&=\sum_{i=1}^m \miop(B_i;Y|B^{i-1})\\
&=\sum_{i=1}^m \entop(B_i|B^{i-1})-\entop(B_i|YB^{i-1})\\
&\ogeq{a}\sum_{i=1}^m \entop(B_i|B^{i-1})-\entop(B_i|Y)\\
&\oeq{b}\entop(\vecB)-\sum_{i=1}^m\entop(B_i|Y)\label{eq:bmd:yrate}
\end{align}
where (a) follows because conditioning does not increase entropy \cite[Theorem~2.6.5]{cover2006elements} and where we used the chain rule for entropy in (b). The expression in the last line of \eqref{eq:bmd:yrate} can be approached as follows. We jointly encode all bit-levels by a single binary code of block length $m\nc$. This idea was introduced in \cite{zehavi1992psk} and is now usually called \emph{bit-interleaved coded modulation} (BICM) \cite{ifabregas2008bit,caire1998bit}. Since our PAS transmitter encodes all bit-levels by a single binary code $\matg=[\mati_k|\matp]$, our transmitter is a BICM encoder. At the receiver, we use a \emph{bit-metric decoder}. 
\subsection{Bit-Metric Decoding}
A soft-demapper calculates for each bit-level $i$ the soft-information
\begin{align}
L_i=\underbrace{\log\frac{P_{B_i}(0)}{P_{B_i}(1)}}_\text{a-priori information}+\underbrace{\log\frac{p_{Y|B_i}(Y|0)}{p_{Y|B_i}(Y|1)}}_\text{channel likelihood}.\label{eq:lvalue}
\end{align}
The distribution $P_{B_i}$ and the conditional density $p_{Y|B_i}$ in \eqref{eq:lvalue} can be calculated as
\begin{align}
&P_{B_i}(b_i)=\sum_{\veca\in\{0,1\}^m\colon a_i=b_i}P_{\vecB}(\veca)\\
&p_{Y|B_i}(y|b_i)=\sum_{\veca\in\{0,1\}^m\colon a_i=b_i}p_{Y|\vecB}(y|\veca)\frac{P_{\vecB}(\veca)}{P_{B_i}(b_i)}.\label{eq:pby}
\end{align}
The soft-information $L_i$ forms a sufficient statistic to estimate bit level $B_i$ from the channel output $Y$ \cite[Sec.~8.2.1]{gallager2013stochastic}, i.e., we have 
%\begin{align}
$\miop(B_i;Y)=\miop(B_i;L_i)$. %\label{eq:bmd:sufficient}
%\end{align}
 A bit-metric decoder uses the soft information $L_1,L_2,\dotsc,L_m$ to estimate the transmitted data and achieves the rate \cite[Theorem~1]{bocherer2014achievable}
\begin{align}
\mathsf{R}_{\text{BMD}}=\entop(\vecB)-\sum_{i=1}^m \entop(B_i|Y).\label{eq:bmd:rate}
\end{align}
Note that \eqref{eq:bmd:rate} is equal to the expression we derived in \eqref{eq:bmd:yrate}.
\begin{remark}
If the bit levels are independent, the rate becomes the ``BICM capacity" \cite[Theorem~1]{martinez2009bit}
\begin{align}
\sum_{i=1}^m \miop(B_i;Y).\label{eq:rbicm}
\end{align}
Using non-uniform distributions for the independent bit-levels results in bit-shaping, which was introduced in \cite{ifabregas2010bit} and is discussed in \cite{bocherer2014achievable}. The authors in \cite[Eq.~(1)]{kayhan2012constellation} call \eqref{eq:rbicm} the ``pragmatic capacity".
\end{remark}
From a coding perspective, bit-metric decoding transforms the channel \eqref{eq:labelchannel} into $m$ parallel binary input channels $p_{L_i|B_i}$, $i=1,2,\dotsc,m$.
%We illustrate this in Fig.~\ref{fig:equivalentchannel} and make the following remarks.
%\begin{figure}
%\footnotesize
%\centering
%\input{diagrams/L_values_model}
%\caption{Coding perspective on bit-metric decoding.}
%\label{fig:equivalentchannel}
%\end{figure}

%\begin{align}
%\begin{split}
%B_1\text{---}&\boxed{P_{L_1|B_1}}\text{---}L_1\\
%B_2\text{---}&\boxed{P_{L_2|B_2}}\text{---}L_2\\
%&\quad\vdots\\
%B_m\text{---}&\boxed{P_{L_m|B_m}}\text{---}L_m
%\end{split}
%\end{align}

\begin{enumerate}
\item The bit channels $p_{L_i|B_i}$ 
%in Fig.~\ref{fig:equivalentchannel} 
are different for different bit levels. This is important in Sec.~\ref{sec:code}, where we optimize the parity matrix $\matp$ for bit-metric decoding.
\item By \eqref{eq:pby}, the distribution of the bits $B_j$, $j\neq i$ influences the channel transition probabilities $p_{L_i|B_i}$ of the $i$th bit-channel, i.e., for each bit level, the other bit-levels act as interference.
\item The rate \eqref{eq:bmd:rate} can take on different values for different labelings of the signal points. We discuss this next. 
\end{enumerate}
\subsection{Optimizing the Labeling}
\label{subsec:labeling}
\begin{table}
\footnotesize
\centering
\caption{Two labelings for the amplitudes of 8-ASK and the resulting signal point labelings}
\label{tab:labelings}
\begin{tabular}{r|cccc}
Amplitude&7&5&3&1\\\hline
natural labeling&00&01&10&11\\
BRGC&00&01&11&10
\end{tabular}
\begin{tabular}{r|cccccccc}
Signal point&-7&-5&-3&-1&1&3&5&7\\\hline
Labeling 1&000&001&010&011&111&110&101&100\\
BRGC&000&001&011&010&110&111&101&100
\end{tabular}
\end{table}
\begin{figure}
%\begin{minipage}[t]{0.48\textwidth}
\centering
\footnotesize
\includegraphics{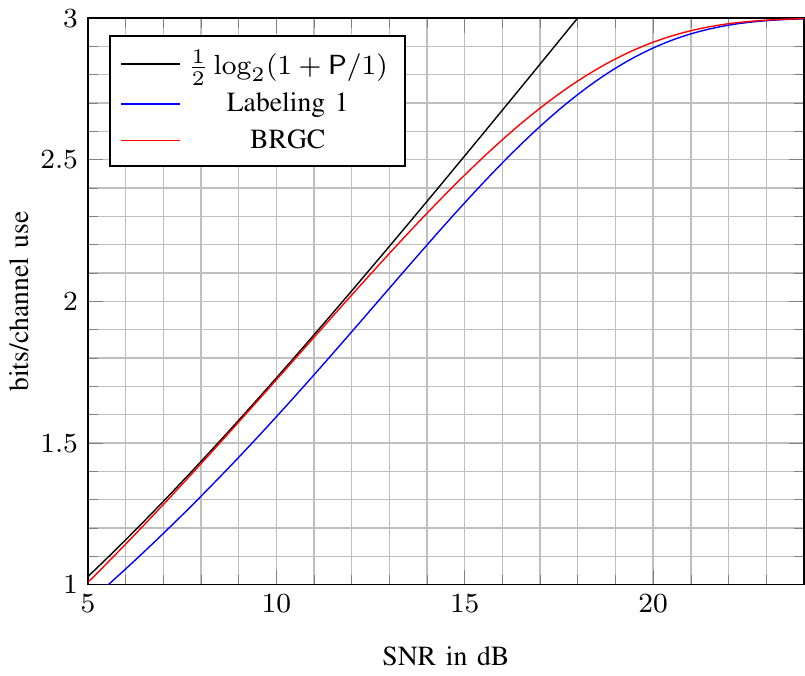}
\caption{Comparison of $\bmdrate$ for the two labelings from Table~\ref{tab:labelings}.}
\label{fig:labelings}
\end{figure}
%\end{minipage}
%\hfill
%\\[0.5cm]
\begin{table}
\footnotesize
\caption{Comparison of SNRs needed by symbol-metric decoding and bit-metric decoding to achieve a certain rate}
\label{tab:bmdgap}
\centering
\begin{tabular}{rrrrr}
\!\!constellation&\!rate&\!SNR($\smdrate$) [dB]&\!SNR($\bmdrate$) [dB]&\!Gap [dB]\\\hline
4-ASK&1&4.8180&4.8313&0.0133\\
8-ASK&2&11.8425&11.8481&0.0056\\
16-ASK&3&18.0911&18.0951&0.0039\\
32-ASK&4&24.1708&24.1742&0.0034\\
64-ASK&5&30.2078&30.2110&0.0032
\end{tabular}
\end{table}

We choose the labeling of the $m-1$ amplitudes; the label of the sign is already defined by construction. We evaluate $\bmdrate$ for $8$-ASK and two different amplitude labelings. Labeling 1 is a \emph{natural labeling} and Labeling 2 is the \emph{Binary Reflected Gray Code} (BRGC), see \cite{gray1953pulse}. The two labelings are listed in Table~\ref{tab:labelings}. We have ordered the amplitudes in descending order so that the actual signal points are labeled from left to right. The resulting labelings of the signal points are also displayed in Table~\ref{tab:labelings}. Note that the natural labeling of the amplitudes does not lead to a natural labeling of the signal points. In contrast, the BRGC labeling of the amplitudes also leads to the BRGC of the signal points. We display in Fig.~\ref{fig:labelings} $\bmdrate$ for Labeling 1 and the BRGC labeling. The BRGC labeling is better than Labeling 1 and very close to the capacity $\capacity(\power)$, which is consistent with the results presented in \cite{bocherer2014achievable}. In Table~\ref{tab:bmdgap}, bit-metric decoding with BRGC is compared to symbol-metric decoding. The SNRs needed to achieve a given rate are listed.
\begin{remark}
The bit-metric decoding gaps in Table~\ref{tab:bmdgap} show how much we can gain by using multilevel coding with multistage decoding or by iteratively exchanging extrinsic information between the soft-demapper and the soft-decoder (BICM-ID, \cite{li1997bit}). This gain is negligible and not worth the increased complexity for the considered scenario. 
\end{remark}

%%%%%%%%%%%%%%%%%%%%%%%%%%%%%%%
%%%%%%%%%%%%%%%%%%%%%%%%%%%%%%%
%%%%%%%%%%%%%%%%%%%%%%%%%%%%%%%
% CODE DESIGN %%%%%%%%%%%%%%%%%
%%%%%%%%%%%%%%%%%%%%%%%%%%%%%%%
%%%%%%%%%%%%%%%%%%%%%%%%%%%%%%%

\section{LDPC Code Design}
\label{sec:code}

In principle, our scheme works for any binary code with a systematic encoder and a decoder that can process the soft-output of the binary demappers. In this section, we discuss the deployment of LDPC codes. 

\subsection{LDPC Codes and Bit-Channels}
%\begin{figure}
%\centering
%\footnotesize
%\input{figures/dvbs2_cmp_mapping.tex}
%\caption{Influence of the bit-mapping on the performance. The DVB-S2 rate 2/3 LDPC code is used with uniform inputs on an 8-ASK constellation. An optimized bit-mapper (blue curve) and a random bit-mapper (red curve) are used.}
%\label{fig:mapping_influence}
%\end{figure}
LDPC codes are linear block codes with a sparse $(n-k)\times n$ check matrix $\mathh$. The matrix $\mathh$ can be represented by a Tanner graph \cite{tanner_graph} consisting of variable nodes and check nodes. The variable node degree of the $i$th coded bit is given by the number of ones in the $i$th column of $\mathh$ and the degree of the $j$th check node is given by the number of ones in the $j$th row of $\mathh$. The variable and check node degrees influence the decoding threshold of the LDPC code \cite{richardson2001design}.

\begin{figure}
%\begin{minipage}[t]{0.48\textwidth}
%\input{figures/dvbs2_cmp_mapping.tex}
\includegraphics{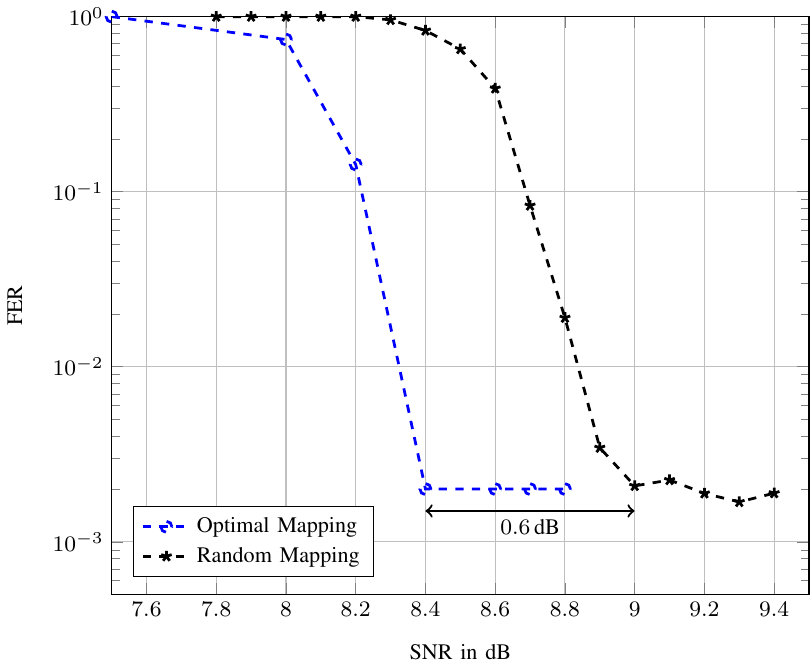}
\caption{Influence of the bit-mapping on the performance. The DVB-S2 rate 2/3 LDPC code is used with uniform inputs on an 8-ASK constellation. An optimized bit-mapper (blue curve) and a random bit-mapper (red curve) are used.}
\label{fig:mapping_influence}
%\end{minipage}
\end{figure}

Good LDPC codes are often irregular, i.e., not all coded bits have the same variable node degree \cite{luby2001improved,richardson2001design}. At the same time, the coded bits are transmitted over different bit-channels. This suggests that the bit-mapper, which decides which coded bit is transmitted over which bit-channel, influences the performance. The example shown in Fig.~\ref{fig:mapping_influence} confirms this. The optimized bit-mapper is 0.6 dB more energy efficient than a randomly chosen bit-mapper.

Bit-mapper optimization was considered, e.g., in \cite{li2005bit,lei2008,gong2011improve,cheng2012exit,hager2014optimized}. An alternative approach is to jointly optimize the node degrees and the bit-mapper. This is done in \cite{zhang_MET_journal} and \cite{steiner_boecherer_2015}. In this work, we consider bit-mapper optimization.

\subsection{Bit-Mapper Optimization for the DVB-S2 Codes}
\label{sec:bitmapper}

We use LDPC codes from the DVB-S2 standard. In Table~\ref{tab:dvbs2_degrees}, we display the variable node degree distributions of the DVB-S2 codes, e.g., the rate 2/3 code has 4320 coded bits of variable node degree 13, and it has 38880, 21599, and 1 bits of degrees 3, 2, and 1, respectively. All codes have four different variable node degrees, which decrease from left to right. By Sec.~\ref{subsec:encoding}, our PAS scheme places the uniformly distributed bits of bit-level $B_1$ at the end of the codeword in the systematic encoding process. For the bit-levels $B_2\dotsb B_m$, we use the following heuristic.
\begin{itemize}
\item A bit interleaver $\pi_\text{b}$ sorts the bit stream by bit-levels, i.e.,
\begin{align}
\vecb(A_1)\dotsc\vecb(A_{\nc})\overset{\pi_\text{b}}{\longrightarrow}\vecB_2\vecB_3\dotsb\vecB_m
\end{align}
where $\vecB_i$ is a string of $\nc$ bits of level $i$.
\item A bit-level interleaver permutes the bit-level strings. We compactly represent the bit-level interleaver by listing the bit-levels in the order in which they should occur in the codeword, e.g., for $m=4$, the bit-level interleaver $(4,2,3)$ is
\begin{align}
\vecB_2\vecB_3\vecB_4\overset{(4,2,3)}{\longrightarrow}\vecB_4\vecB_2\vecB_3.\label{eq:blexample}
\end{align}  
\end{itemize}
We keep the bit interleaver fixed and represent the complete bit-mapping by appending a $1$ to the bit-level interleaver, e.g., $\bitm=(4,2,3,1)$ stands for the composition of $\pi_\text{b}$ with bit-level interleaver $(4,2,3)$. We optimize the bit-level interleaver. There are $(m-1)!$ possibilities, among which we choose the one with the best error performance. The largest considered constellation is 64-ASK, for which we need to choose among $(6-1)!=120$ bit level interleavers. This optimization is still feasible. We display in Table~\ref{tab:interleaver} the optimized bit-mappers for uniform inputs and rate $(m-1)/m$ codes.
%\begin{table}
%\caption{Optimal bit-level interleaver for uniform inputs and DVB-S2 codes}
%\label{tab:interleaver}
%\centering
%\begin{tabular}{rcr}
% \toprule
% constellation & code rate & bit-mapper $\pi$\\
% \midrule
% 8ASK &2/3 &$(3,2,1)$\\
% 16ASK &3/4 &$(3,4,2,1)$\\
% 32ASK &4/5 &$(3,2,5,4,1)$\\
% 64ASK &5/6 &$(4,2,5,3,6,1)$\\
% \bottomrule
%\end{tabular}
%	\end{table}
\begin{table}
\footnotesize
\begin{minipage}[t]{0.48\textwidth}
\centering
\caption{Variable Node Degree Distributions of DVB-S2 Codes}
\begin{tabular}{llllllll}
  \toprule
   & \multicolumn{7}{c}{variable node degrees}\\
  \cmidrule{2-8}
  rate   & 13 & 12 & 11 & 4 & 3 & 2 & 1\\
  \midrule
   2/3 & 4320 & & & & 38880 & 21599 & 1\\
   3/4 & & 5400 &  & & 43200 & 16199 & 1\\
   4/5 &        &       & 6480 &  & 45360 & 12959 & 1\\
   5/6 & 5400   &       &   &   & 48600 & 10799 & 1\\
   9/10&  & & & 6480 & 51840 & 6479 & 1\\
   \bottomrule
\end{tabular}
\label{tab:dvbs2_degrees}
\end{minipage}
%\hfill
\\[0.5cm]
\begin{minipage}[t]{0.48\textwidth}
\centering
\caption{Optimal bit-level interleaver for uniform inputs and DVB-S2 codes}
\begin{tabular}{rcr}
 \toprule
 constellation & code rate & bit-mapper $\pi$\\
 \midrule
 8ASK &2/3 &$(3,2,1)$\\
 16ASK &3/4 &$(3,4,2,1)$\\
 32ASK &4/5 &$(3,2,5,4,1)$\\
 64ASK &5/6 &$(4,2,5,3,6,1)$\\
 \bottomrule
\end{tabular}
\label{tab:interleaver}
\end{minipage}
\end{table}
%%%%%%%%%%%%%%%%%%%%%%%%%%%%%%%
%%%%%%%%%%%%%%%%%%%%%%%%%%%%%%%
%%%%%%%%%%%%%%%%%%%%%%%%%%%%%%%
% RATE ADAPTION %%%%%%%%%%%%%%%
%%%%%%%%%%%%%%%%%%%%%%%%%%%%%%%
%%%%%%%%%%%%%%%%%%%%%%%%%%%%%%%

\section{Rate Adaption}
\label{sec:adaption}

\subsection{Practical Operating Points}
\label{subsec:practicalops}
\begin{figure}
\centering
\footnotesize
%\input{figures/adaption}
%\begin{minipage}[t]{0.48\textwidth}
\includegraphics{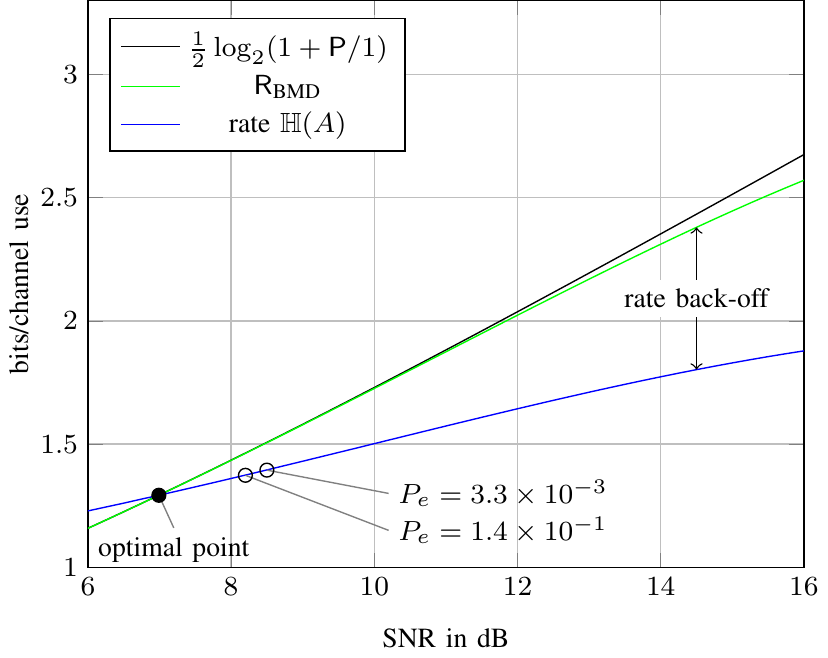}
\caption{Operating points for 8-ASK and rate $2/3$ codes. The optimal point is where the rate curve $R(\power)$ crosses the achievable rate curve $\bmdrate(\power)$. By backing off from the optimal point along the rate curve, the rate back-off (given by the distance between the rate curve and the achievable rate curve) increases. For the rate 2/3 DVB-S2 LDPC code, the FER $P_e$ is decreased from $\num{1.1e-1}$ to $\num{2.8e-3}$ by increasing the rate back-off.}
\label{fig:backoff}
\end{figure}

%\begin{figure}
%\footnotesize
%\input{figures/uncertainity}
%\caption{Operating points for 8-ASK and rate $2/3$ codes. The optimal point is where the rate curve $R(\power)$ crosses the achievable rate curve $\bmdrate(\power)$. By backing off from the optimal point along the rate curve, the rate back-off (given by the distance between the rate curve and the achievable rate curve) increases. For the rate 2/3 DVB-S2 LDPC code, the FER $P_e$ is decreased from $\num{1.1e-1}$ to $\num{2.8e-3}$ by increasing the rate back-off.}
%\label{fig:backoff}
%\end{figure}
We use PAS at the transmitter and bit-metric decoding at the receiver. The transmission rate and the achievable rate are given by the respective
\begin{align}
R=\entop(A)+\gamma,\qquad\bmdrate&=\entop(\vecB)-\sum_{i=1}^m H(B_i|L_i)
\end{align}
where $\vecB=b(S)\vecb(A)$. For each transmit power $\power$, we choose the amplitude distribution $P_A$ and the constellation scaling $\Delta$ to maximize the achievable rate. In this way, we obtain a transmission rate curve $R$ and an achievable rate curve $\bmdrate$. In analogy to our discussion in Sec.~\ref{subsec:pas:ops}, the optimal operating point for our scheme is where $R$ crosses $\bmdrate$. We illustrate this in Fig.~\ref{fig:backoff} for 8-ASK and code rate $c=2/3$ ($\gamma=0$). When we use practical codes of finite block length $\nc$, we must back off from the optimal operating point and tolerate a positive FER. We back off along the transmission rate curve $R$. As we increase the transmission power $\power$, the \emph{rate back-off}
\begin{align}
\bmdrate-R
\end{align}
increases and we expect that the error probability decreases. This intuition is confirmed by the practical operating points of the 2/3 DVB-S2 LDPC code shown in Fig.~\ref{fig:backoff}.

\subsection{Transmission Rate Adaption}

Suppose we achieve a transmission rate $R^\circ$ at a FER $P_e$ by the procedure described in Sec.~\ref{subsec:practicalops}. Suppose further that this \emph{reference operating point} is achieved by amplitude distribution $P_A$ and constellation scaling $\Delta$. With the same code, we now want to transmit at a rate $\tilde{R}\neq R^\circ$ and achieve the same error probability.

Recall that for some optimized $\nu$ (see Sec.~\ref{subsec:optimizedinput}), the input distribution $P_{\heux}$ is given by
\begin{align}
P_{\heux}(x)=\frac{e^{-\nu |x|^2}}{\sum_{x'\in\setx} e^{-\nu |x'|^2}},\quad x\in\setx.
\end{align}
By \eqref{eq:pas:pa}, we can write the amplitude distribution $P_A$ of our reference point as
\begin{align}
P_{A}(a)=2P_{\heux}(a),\quad a\in\seta.
\end{align}
Now define
\begin{align}
P_{A^\lambda}(a)=\frac{P_A(a)e^{\lambda a^2}}{\sum_{\tilde{a}\in\seta}P_A(\tilde{a})e^{\lambda \tilde{a}^2}},\quad a\in\seta.\label{eq:mbscaling}
\end{align}
The distribution $P_{A^\lambda}$ has the following properties:
\begin{itemize}
\item $P_A$ and $P_{A^\lambda}$ are Maxwell-Boltzmann distributions. In particular, the set of Maxwell-Boltzmann distributions is closed under the mapping $P_A\mapsto P_{A^\lambda}$ as defined by \eqref{eq:mbscaling}.
\item For $\lambda=0$, we have $P_{A^\lambda}=P_A$.
\item For $\lambda\to \nu$, $P_A$ approaches the uniform distribution on $\seta$ whose entropy is $\log_2|\seta|=m-1$.
\item For $\lambda\to-\infty$, $P_{A^\lambda}$ approaches the distribution that chooses the smallest amplitude with probability $1$. The resulting entropy is zero.
\end{itemize}
From these properties, we see that we can use $P_{A^\lambda}$ to adapt the transmission rate. The range of feasible rates is
\begin{align}
\gamma\leq \entop(A^\lambda)+\gamma\leq m-1+\gamma.
\end{align}
For $\gamma=0$, the rate is between $0$ and $m-1$. To transmit at a feasible rate $R$, we proceed as follows.

\begin{mdframed}
\begin{itemize}
\item Choose the amplitude distribution such that the desired transmission rate is achieved, i.e., choose
\begin{align}
P_{\tilde{A}}=P_{A^\lambda}\colon \entop(A^\lambda)+\gamma=\tilde{R}.
\end{align}
\item Choose the constellation scaling $\tilde{\Delta}$ such that the resulting error probability $\tilde{P}_e$ is the same as for the reference point, i.e., choose
\begin{align}
\tilde{P}_e\overset{!}{=}P_e.
\end{align}
\end{itemize}
\end{mdframed}

\subsection{Adaption for Universal Codes}
\label{sec:universal}
The procedure described in the previous section requires to search for the right constellation scaling $\Delta$ by repeatedly performing Monte Carlo simulations. We now show how this can be simplified significantly for universal codes. We say a code is \emph{universal}, if $P_e$ depends only on the rate back-off $\bmdrate-R$. Suppose for $P_A$ and $\Delta$ of our reference point, the achievable rate evaluates to $\bmdrate^\circ$. We can now adapt the constellation scaling as follows.
\begin{mdframed}
\begin{itemize}
\item Choose the constellation scaling $\tilde{\Delta}$ such that
\begin{align}
\bmdrate(\tilde{\Delta},P_{\tilde{A}})-\tilde{R}=\bmdrate^\circ-R^\circ.\label{eq:adaption strategy}
\end{align}
\end{itemize}
\end{mdframed}
\begin{remark}
For PAS and bit-metric decoding, we can write the rate back-off as
\begin{align}
\bmdrate-R=1-\gamma-\sum_{i=1}^m H(B_i|L_i).
\end{align}
The term $1-\gamma$ is constant, so $P_e$ is determined by the sum of the conditional entropies $\sum_{i=1}^m H(B_i|L_i)$.
\end{remark}
\begin{remark}
For a uniformly distributed binary input $B$ and a rate $c$ code, the transmission rate is $R=c$ and the achievable rate is $R^*=\miop(B;Y)$. The rate back-off becomes 
\begin{align}
R^*-R=\miop(B;Y)-c.\label{eq:universal:mi}
\end{align}
Since $c$ is constant, this shows that $P_e$ is determined by the mutual information $\miop(X;Y)$. This property was observed for practical LDPC codes in \cite{franceschini2006performance}.
\end{remark}

%\end{minipage}
%\hfill
%\\[0.5cm]
%\begin{minipage}[t]{0.48\textwidth}
\begin{figure}
\centering
\includegraphics{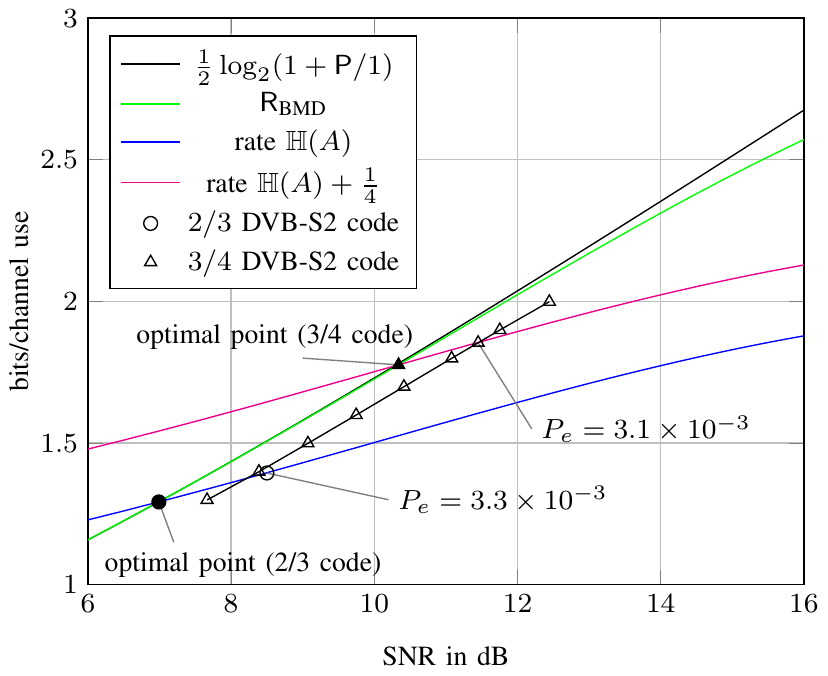}
\caption{Transmission rate adaption for 8-ASK. The rate 3/4 DVB-S2 code is used. The triangles mark the operating points calculated under the assumption that the code is universal. The reference operating point is the triangle on the rate $\entop(A)+\frac{1}{4}$ curve. In Table~\ref{tab:8ask}, we display the FER of the adapted operating points. For comparison, we display an operating point of the 2/3 DVB-S2 code.}
\label{fig:adaption}
%\end{minipage}
\end{figure}

\section{Numerical Results}
\label{sec:numerical}
%%%%%%%%%%%%
%% NUMERICAL
%%%%%%%%%%%%
 \pgfplotstableset{
%    highlight/.append style={
%        postproc cell content/.append code={
%                \pgfkeysalso{@cell content=\textbf{##1}}%
%        },
%    },
    every first row/.style={before row=\toprule,after row=\midrule},
%rate
columns/0/.style={
column name=Rate,
fixed,fixed zerofill,
precision=2},
%snr
columns/1/.style={
column name=SNR [dB],
fixed,fixed zerofill,
precision=2},
%gap
columns/2/.style={
column name=Gap [dB],
fixed,fixed zerofill,
precision=2},
%FER
columns/3/.style={
column name=FER,
sci,sci zerofill,
precision=1},
%dev
columns/4/.style={
column name=95\% CI,
sci,sci zerofill,
precision=1,
postproc cell content/.append style={
/pgfplots/table/@cell content/.add={$\pm}{$},}
},
}
%
%
% 64ASK
%
%
%{\setstretch{1.0}
\begin{table}
%\footnotesize
\centering
%\begin{minipage}{0.48\textwidth}
%\parbox{0.48\textwidth}{
\caption{$64$-ASK, 9/10 DVB-S2, $\bitm=(4,2,5,3,6,1)$}
\label{tab:64ask}
\vspace{-0.15cm}
\pgfplotstabletypeset[
]{
5.0913   31.8006    1.1517    0.0041    0.0021
4.9849   31.0934    1.0856    0.0080    0.0040
 4.8853   30.4589    1.0516    0.0068    0.0034
    4.7853   29.8372    1.0327    0.0073    0.0037
    4.6869   29.2334    1.0217    0.0064    0.0032
    4.5869   28.6237    1.0156    0.0038    0.0019
    4.4860   28.0103    1.0104    0.0062    0.0031
    4.3878   27.4140    1.0069    0.0076    0.0038
    4.2870   26.8017    1.0026    0.0096    0.0048
    4.1894   26.2081    0.9988    0.0101    0.0051
    4.0900   25.6054    0.9962    0.0102    0.0051
    3.9899   24.9966    0.9922    0.0100    0.0050
}
%\end{table}
%
%
% 32ASK
%
%
%\begin{table}
\vspace{0.2cm}
\caption{$32$-ASK, 5/6 DVB-S2, $\bitm=(4,5,2,3,1)$}
\label{tab:32ask}
\vspace{-0.15cm}
\pgfplotstabletypeset{
3.6222   22.5997    0.8205    0.0021    0.0011
 3.9930   25.0929    1.0699    0.0033    0.0017
    3.8932   24.3598    0.9400    0.0018    0.0009
    3.7933   23.6872    0.8720    0.0028    0.0014
    3.6937   23.0474    0.8353    0.0015    0.0008
    3.5935   22.4209    0.8157    0.0016    0.0008
    3.4935   21.8045    0.8058    0.0015    0.0008
    3.3939   21.1949    0.8010    0.0013    0.0006
    3.2938   20.5841    0.7987    0.0008    0.0004
    3.1942   19.9767    0.7978    0.0007    0.0004
    3.0951   19.3714    0.7971    0.0005    0.0003
    2.9953   18.7613    0.7967    0.0007    0.0004
}
%}
%\end{minipage}
%\end{table}
%
%
% 16ASK
%
%
%\begin{table}
%\hfill
%\begin{minipage}{0.48\textwidth}
\vspace{0.2cm}
\caption{$16$-ASK, 5/6 DVB-S2, $\bitm=(4,3,2,1)$}
\label{tab:16ask}
\vspace{-0.15cm}
\pgfplotstabletypeset{
2.9573   18.3997    0.6673    0.0195    0.0098
   2.9973   18.6631    0.6863    0.0296    0.0143
    2.8972   18.0103    0.6466    0.0231    0.0116
    2.7975   17.3785    0.6264    0.0195    0.0098
    2.6973   16.7521    0.6169    0.0156    0.0078
    2.5973   16.1305    0.6135    0.0133    0.0067
    2.4978   15.5126    0.6128    0.0072    0.0036
    2.3978   14.8910    0.6142    0.0035    0.0018
    2.2982   14.2699    0.6166    0.0025    0.0013
    2.1978   13.6401    0.6196    0.0019    0.0009
    2.0978   13.0094    0.6229    0.0013    0.0007
    1.9980   12.3746    0.6268    0.0005    0.0002
}
%\end{table}
%
%
% 8ASK
%
%
%\begin{table}
\vspace{0.2cm}
\caption{$8$-ASK, 3/4 DVB-S2, $\bitm=(3,2,1)$}
\label{tab:8ask}
\vspace{-0.15cm}
\pgfplotstabletypeset{
1.8543   11.4500    0.6316    0.0015    0.0008
 1.9990   12.4446    0.6902    0.0038    0.0019
    1.8991   11.7527    0.6431    0.0021    0.0011
    1.7991   11.0798    0.6226    0.0012    0.0006
    1.6990   10.4136    0.6173    0.0023    0.0012
    1.5991    9.7465    0.6201    0.0012    0.0006
    1.4992    9.0730    0.6278    0.0015    0.0007
    1.3991    8.3892    0.6397    0.0012    0.0006
    1.2994    7.6933    0.6540    0.0022    0.0011
}
%\end{table}
%
%
% 4ASK
%
%
%\begin{table}
\vspace{0.2cm}
\caption{$4$-ASK, 2/3 DVB-S2, $\bitm=(2,1)$}
\label{tab:4ask}
\vspace{-0.15cm}
\pgfplotstabletypeset{
1.1313    6.6999    0.9035    0.0052    0.0026
1.1998    7.2554    0.9448    0.0026    0.0013
    1.0997    6.4498    0.8952    0.0064    0.0033
    0.9998    5.6649    0.8956    0.0136    0.0068
%    0.8998    4.8702    0.9237    0.0481    0.0237
%    0.7998    4.0478    0.9720    0.2917    0.1050
}
%\end{minipage}
\vspace{-0.5cm}
\end{table}
%}
%%%%%%%%%%%%
%% /NUMERICAL
%%%%%%%%%%%%

We assess the performance of our scheme by Monte Carlo simulation. For each block, $\kc$ uniformly distributed data bits are transmitted in $\nc$ channel uses. $\gamma\nc$ of the data bits $U^{\kc}=U_1U_2\dotsb U_{\kc}$ are used for bit-level 1 (see Fig.~\ref{fig:encoderAdapted}). The remaining $\kc-\gamma\nc$ data bits are transformed by a CCDM matcher (see Sec.~\ref{sec:dm}) into a sequence of $\nc$ amplitudes. Encoding is then done according to Fig.~\ref{fig:encoderAdapted}. At the receiver, bit-metric decoding is performed (see Sec.~\ref{sec:bmd}) and the amplitude estimates are transformed back into $\kc-\gamma\nc$ data bit estimates by a CCDM dematcher, see \cite{schulte2015constant}. The dematcher output together with the other $\gamma\nc$ data bit estimates form the data estimate $\hat{U}^{\kc}=\hat{U_1}\hat{U_2}\dotsb\hat{U_{\kc}}$. We estimate the end-to-end FER~$\Pr\{\hat{U}^{\kc}\neq U^{\kc}\}$. For each FER estimate, we also provide the corresponding 95\% confidence interval (CI). The spectral efficiency is 
%\begin{align}
$R=\frac{\kc}{\nc}$.
%\end{align}
We use the DVB-S2 LDPC codes and we decode with 100 iterations. We optimize the bit-mappers using the heuristic proposed in Sec.~\ref{sec:bitmapper}. For each considered (constellation, code rate) mode, we first determine the practical operating point with $\text{FER}\approx\num{1e-3}$ following the procedure described in Sec.~\ref{subsec:practicalops}. We then use the rate adaption \eqref{eq:adaption strategy} to operate the same (constellation, code rate) mode over a range of SNRs and spectral efficiencies. The results are displayed in Table~\ref{tab:64ask}--\ref{tab:4ask}. In the first line of each table, the practical operating point is displayed and separated from the rest of the displayed values by a horizontal line. For 8-ASK, we display the resulting operating points in Fig.~\ref{fig:adaption}.
\subsubsection{Spectral Efficiency} In Table~\ref{tab:64ask}---\ref{tab:4ask}, we display the SNR gap of our scheme to capacity $\capacity(\power)=\frac{1}{2}\log_2(1+\power/1)$. Over a range of 1 bit/s/Hz to 5 bit/s/Hz, our scheme operates within 1 dB of capacity. The rate 5/6 code with 16-ASK has a gap of only 0.61 to 0.69 dB, while the rate 9/10 code with 64-ASK has a gap of 1 dB.  This indicates that the spectral efficiency can be improved further by using optimized codes. The work \cite{steiner_boecherer_2015} confirms this.
\subsubsection{Universality} The results in Table~\ref{tab:64ask}---\ref{tab:4ask} show that the DVB-S2 codes are universal in the sense of Sec.~\ref{sec:universal}, since over the whole considered range, the resulting FER is within the waterfall region. Some configurations show a ``more universal" behavior than others, e.g., the FER of the rate 3/4 code with 8-ASK (Table~\ref{tab:8ask}) is almost the same for the spectral efficiencies from 1.3 to 2.0 bits/s/Hz, while for the rate 5/6 code with 16-ASK, the FER changes by almost two orders of magnitude. Note that the listed operating points were found by \eqref{eq:adaption strategy}. The operating points that result in a too low  (too high) FER could be corrected by decreasing (increasing) the transmission power. 
\subsubsection{Rate} Since all considered codes have the same block length of 64800 bits, the number of channel uses $\nc$ gets smaller for larger constellations. As discussed in Sec.~\ref{sec:dm}, the CCDM matcher that we use to emulate the amplitude source $P_A$ has a rate close to $\entop(A)$ when the output length $\nc$ is large. For the smaller constellations 4, 8 and 16-ASK, the effective rate is equal to $\entop(A)$ with a precision of two decimal places. For the larger constellations 32 and 64-ASK, we observe a slight decrease of the effective rate, e.g., instead of the target rate $5.00$ we observe an effective rate $R=4.98$. See also \cite{schulte2015constant} for a discussion of this phenomenon. We calculate the SNR gap with respect to the effective rate.
\subsubsection{Error Floor} The DVB-S2 codes are designed to get the FER down to a reasonable value and an outer code then lowers the error probability further \cite{etsi2009dvb}. Since the CCDM dematcher performs a non-linear transformation, one corrupted symbol at the dematcher input can lead to various corrupted bit-errors at the dematcher output, so in-block error propagation can occur. In our scheme, an outer encoder should therefore be placed between the matcher and the inner encoder and the outer decoder should be placed between the inner decoder and the dematcher. To preserve the amplitude distribution imposed by the matcher, a systematic outer encoder has to be used.

%%%%%%%%%%%%%%%%%%%%%%%%%%%%%%%
%%%%%%%%%%%%%%%%%%%%%%%%%%%%%%%
%%%%%%%%%%%%%%%%%%%%%%%%%%%%%%%
% CONCLUSIONS %%%%%%%%%%%%%%%%%
%%%%%%%%%%%%%%%%%%%%%%%%%%%%%%%
%%%%%%%%%%%%%%%%%%%%%%%%%%%%%%%

\section{Conclusions}
\label{sec:conclusions}
We proposed a practical rate-matched coded modulation scheme that adapts its transmission rate to the SNR. At the transmitter, a distribution matcher is concatenated with a systematic encoder of a binary LDPC code. At the receiver, bit-metric decoding is used. No iterative demapping is required. The reported numerical results show that on the complex baseband AWGN channel, any spectral efficiency between 2--10 bit/s/Hz can be achieved within 1 dB of capacity $\log_2(1+\power/1)$ by using the off-the-shelf DVB-S2 LDPC codes of rate 2/3, 3/4, 4/5 and 9/10 together with QAM constellations with up to 4096 signal points. Future work should investigate rate-matched coded modulation for short block lengths and for multiple-input multiple-output channels.

%\appendices

%%%%%%%%%%%%%%%%%%%%%%%%%%%%%%%
%%%%%%%%%%%%%%%%%%%%%%%%%%%%%%%
%%%%%%%%%%%%%%%%%%%%%%%%%%%%%%%
% APPENDIX %%%%%%%%%%%%%%%%%%%%
%%%%%%%%%%%%%%%%%%%%%%%%%%%%%%%
%%%%%%%%%%%%%%%%%%%%%%%%%%%%%%%

% \section{Coded Modulation Converse}
%\label{app:converse}
%Let $U^k$ be the independent and uniformly distributed source bits and let the codeword $X^n$ be a deterministic function of $U^k$. The transmission rate is $R=k/n$ [bits/channel use]. The codeword is transmitted over a memoryless channel $P_{Y|X}$. Let $\hat{U}^k$ be the decoder output. The decoding error is $P_e=\Pr\{\hat{U}^n\neq U^n\}$. We have
%\begin{align*}
%\entop_2(P_e)+P_e\log_2(2^k)&\ogeq{a} \entop(U^k|\hat{U}^k)\\
%&=\entop(U^k)-\miop(U^k;\hat{U}^k)\\
%&\ogeq{b} k-\miop(X^n;Y^n)\\
%&= k-[\entop(Y^n)-\entop(Y^n|X^n)]\\
%&\oeq{c} k-\Bigl[\entop(Y^n)-\sum_{i=1}^n\entop(Y_i|X_i)\Bigr]\\
%&\ogeq{d} k-\sum_{i=1}^n\bigl[\entop(Y_i)-\entop(Y_i|X_i)\bigr]\\
%&= k-\sum_{i=1}^n \miop(X_i;Y_i)
%\end{align*}
%where (a) follows by Fano's inequality \cite[Theorem~2.10.1]{cover2006elements}, (b) by the data-processing inequality \cite[Theorem~2.8.1]{cover2006elements}, (c) follows because the channel is memoryless and (d) follows by the independence bound on entropy \cite[Theorem~2.6.6]{cover2006elements}. Dividing by $n$, we have
%\begin{align*}
%R-\frac{\sum_{i=1}^n \miop(X_i;Y_i)}{n}\leq \frac{\entop_2(P_e)}{n}+P_eR\leq \entop_2(P_e)+P_eR.
%\end{align*}
%That is, if $R>\frac{\sum_{i=1}^n \miop(X_i;Y_i)}{n}$ then $P_e$ is bounded away from zero.

%\newcommand{\CLASSINPUTbaselinestretch}{1}
%\begin{spacing}{1.24}
\bibliographystyle{IEEEtran}
%\normalsize
\bibliography{IEEEabrv,confs-jrnls,references}

\begin{thebibliography}{10}
\providecommand{\url}[1]{#1}
\csname url@samestyle\endcsname
\providecommand{\newblock}{\relax}
\providecommand{\bibinfo}[2]{#2}
\providecommand{\BIBentrySTDinterwordspacing}{\spaceskip=0pt\relax}
\providecommand{\BIBentryALTinterwordstretchfactor}{4}
\providecommand{\BIBentryALTinterwordspacing}{\spaceskip=\fontdimen2\font plus
\BIBentryALTinterwordstretchfactor\fontdimen3\font minus
  \fontdimen4\font\relax}
\providecommand{\BIBforeignlanguage}[2]{{%
\expandafter\ifx\csname l@#1\endcsname\relax
\typeout{** WARNING: IEEEtran.bst: No hyphenation pattern has been}%
\typeout{** loaded for the language `#1'. Using the pattern for}%
\typeout{** the default language instead.}%
\else
\language=\csname l@#1\endcsname
\fi
#2}}
\providecommand{\BIBdecl}{\relax}
\BIBdecl

\bibitem{etsi2009dvb}
\emph{Digital Video Broadcasting {(DVB)}; 2nd Generation Framing Structure,
  Channel Coding and Modulation Systems for Broadcasting, Interactive Services,
  News Gathering and Other Broadband Satellite Applications {(DVB-S2)}},
  European Telecommun. Standards Inst. (ETSI) Std. EN 302 307, Rev. 1.2.1,
  2009.

\bibitem{raphaeli2004constellation}
D.~Raphaeli and A.~Gurevitz, ``Constellation shaping for pragmatic turbo-coded
  modulation with high spectral efficiency,'' \emph{{IEEE} Trans. Commun.},
  vol.~52, no.~3, pp. 341--345, 2004.

\bibitem{yankov2014rate}
M.~Yankov, S.~Forchhammer, K.~J. Larsen, and L.~P. Christensen, ``Rate-adaptive
  constellation shaping for near-capacity achieving turbo coded {BICM},'' in
  \emph{Proc. IEEE Int. Conf. Commun. (ICC)}, 2014, pp. 2112--2117.

\bibitem{etsi2014dvb}
\emph{Digital Video Broadcasting {(DVB)}; Second generation framing structure,
  channel coding and modulation systems for Broadcasting, Interactive Services,
  News Gathering and other broadband satellite applications; Part 2: {DVB-S2}
  Extensions {(DVB-S2X)}}, European Telecommun. Standards Inst. (ETSI) Std. EN
  302 307-2, Rev. 1.1.1, 2014.

\bibitem{hagenauer1988rate}
J.~Hagenauer, ``Rate-compatible punctured convolutional codes ({RCPC} codes)
  and their applications,'' \emph{{IEEE} Trans. Commun.}, vol.~36, no.~4, pp.
  389--400, 1988.

\bibitem{li2002rate}
J.~Li and K.~R. Narayanan, ``Rate-compatible low density parity check codes for
  capacity-approaching {ARQ} schemes in packet data communications.'' in
  \emph{Int. Conf. Commun., Internet, Inf. Technol. (CIIT)}, 2002, pp.
  201--206.

\bibitem{ha2004rate}
J.~Ha, J.~Kim, and S.~W. McLaughlin, ``Rate-compatible puncturing of
  low-density parity-check codes,'' \emph{{IEEE} Trans. Inf. Theory}, vol.~50,
  no.~11, pp. 2824--2836, 2004.

\bibitem{nguyen2012design}
T.~V. Nguyen, A.~Nosratinia, and D.~Divsalar, ``The design of rate-compatible
  protograph {LDPC} codes,'' \emph{{IEEE} Trans. Commun.}, vol.~60, no.~10, pp.
  2841--2850, 2012.

\bibitem{chen2015protograph}
T.-Y. Chen, K.~Vakilinia, D.~Divsalar, and R.~D. Wesel, ``Protograph-based
  raptor-like {LDPC} codes,'' \emph{{IEEE} Trans. Commun.}, 2015, to appear.

\bibitem{ifabregas2008bit}
A.~Guill\'en~i F\`abregas, A.~Martinez, and G.~Caire, ``Bit-interleaved coded
  modulation,'' \emph{Found. Trends Comm. Inf. Theory}, vol.~5, no. 1--2, pp.
  1--153, 2008.

\bibitem{martinez2009bit}
A.~Martinez, A.~Guill{\'e}n~i F{\`a}bregas, G.~Caire, and F.~Willems,
  ``Bit-interleaved coded modulation revisited: A mismatched decoding
  perspective,'' \emph{{IEEE} Trans. Inf. Theory}, vol.~55, no.~6, pp.
  2756--2765, 2009.

\bibitem{bocherer2014achievable}
\BIBentryALTinterwordspacing
G.~B\"ocherer, ``Achievable rates for shaped bit-metric decoding,'' \emph{arXiv
  preprint}, 2015. [Online]. Available: \url{http://arxiv.org/abs/1410.8075}
\BIBentrySTDinterwordspacing

\bibitem{forney1984efficient}
J.~Forney, G., R.~Gallager, G.~Lang, F.~Longstaff, and S.~Qureshi, ``Efficient
  modulation for band-limited channels,'' \emph{{IEEE} J. Sel. Areas Commun.},
  vol.~2, no.~5, pp. 632--647, 1984.

\bibitem{ling2014achieving}
C.~Ling and J.-C. Belfiore, ``Achieving {AWGN} channel capacity with lattice
  {G}aussian coding,'' \emph{{IEEE} Trans. Inf. Theory}, vol.~60, no.~10, pp.
  5918--5929, 2014.

\bibitem{mondelli2014achieve}
M.~Mondelli, S.~H. Hassani, and R.~Urbanke, ``How to achieve the capacity of
  asymmetric channels,'' in \emph{Proc. Allerton Conf. Commun., Contr.,
  Comput.}, 2014, pp. 789--796.

\bibitem{gallager1968information}
R.~G. Gallager, \emph{Information Theory and Reliable Communication}.\hskip 1em
  plus 0.5em minus 0.4em\relax John Wiley \& Sons, Inc., 1968.

\bibitem{schreckenbach2005signal}
F.~Schreckenbach and P.~Henkel, ``Signal shaping using non-unique symbol
  mappings,'' in \emph{Proc. Allerton Conf. Commun., Contr., Comput.}, Sep.
  2005.

\bibitem{bocherer2013optimal}
G.~B{\"o}cherer, ``Optimal non-uniform mapping for probabilistic shaping,'' in
  \emph{Proc. Int. ITG Conf. Source Channel Coding (SCC)}, 2013.

\bibitem{bocherer2014optimal}
\BIBentryALTinterwordspacing
G.~B\"ocherer and B.~C. Geiger, ``Optimal quantization for distribution
  synthesis,'' \emph{arXiv preprint}, 2014. [Online]. Available:
  \url{http://arxiv.org/abs/1307.6843}
\BIBentrySTDinterwordspacing

\bibitem{forney1992trellis}
J.~Forney, G.~D., ``Trellis shaping,'' \emph{{IEEE} Trans. Inf. Theory},
  vol.~38, no.~2, pp. 281--300, 1992.

\bibitem{tretter2002constellation}
S.~A. Tretter, \emph{Constellation Shaping, Nonlinear Precoding, and Trellis
  Coding for Voiceband Telephone Channel Modems with Emphasis on ITU-T
  Recommendation V. 34}.\hskip 1em plus 0.5em minus 0.4em\relax Kluwer Academic
  Publishers, 2002.

\bibitem{fischer2002precoding}
R.~F.~H. Fischer, \emph{Precoding and Signal Shaping for Digital
  Transmission}.\hskip 1em plus 0.5em minus 0.4em\relax John Wiley \& Sons,
  Inc., 2002.

\bibitem{fischer1998combination}
R.~Fischer, J.~Huber, and U.~Wachsmann, ``On the combination of multilevel
  coding and signal shaping,'' in \emph{Proc. Int. ITG Conf. Source Channel
  Coding (SCC)}.\hskip 1em plus 0.5em minus 0.4em\relax Citeseer, 1998.

\bibitem{wachsmann1999multilevel}
U.~Wachsmann, R.~F.~H. Fischer, and J.~B. Huber, ``Multilevel codes:
  theoretical concepts and practical design rules,'' \emph{{IEEE} Trans. Inf.
  Theory}, vol.~45, no.~5, pp. 1361--1391, 1999.

\bibitem{zehavi1992psk}
E.~Zehavi, ``8-{PSK} trellis codes for a {R}ayleigh channel,'' \emph{{IEEE}
  Trans. Commun.}, vol.~40, no.~5, pp. 873--884, 1992.

\bibitem{caire1998bit}
G.~Caire, G.~Taricco, and E.~Biglieri, ``Bit-interleaved coded modulation,''
  \emph{{IEEE} Trans. Inf. Theory}, vol.~44, no.~3, pp. 927--946, 1998.

\bibitem{smith2012pragmatic}
B.~P. Smith and F.~R. Kschischang, ``A pragmatic coded modulation scheme for
  high-spectral-efficiency fiber-optic communications,'' \emph{J. Lightw.
  Technol.}, vol.~30, no.~13, pp. 2047--2053, 2012.

\bibitem{kaimalettu2007constellation}
S.~Kaimalettu, A.~Thangaraj, M.~Bloch, and S.~McLaughlin, ``Constellation
  shaping using {LDPC} codes,'' in \emph{Proc. IEEE Int. Symp. Inf. Theory
  (ISIT)}.\hskip 1em plus 0.5em minus 0.4em\relax IEEE, 2007, pp. 2366--2370.

\bibitem{khandani1993shaping}
A.~K. Khandani and P.~Kabal, ``Shaping multidimensional signal spaces. {I.}
  optimum shaping, shell mapping,'' \emph{{IEEE} Trans. Inf. Theory}, vol.~39,
  no.~6, pp. 1799--1808, 1993.

\bibitem{kschischang1994optimal}
F.~R. Kschischang and S.~Pasupathy, ``Optimal shaping properties of the
  truncated polydisc,'' \emph{{IEEE} Trans. Inf. Theory}, vol.~40, no.~3, pp.
  892--903, 1994.

\bibitem{itut1998modem}
\BIBentryALTinterwordspacing
{ITU-T Recommendation V.34}, ``A modem operating at data signalling rates of up
  to 33 600 bit/s for use on the general switched telephone network and on
  leased point-to-point 2-wire elephone-type circuits,'' Feb. 1998. [Online].
  Available: \url{http://www.itu.int/rec/T-REC-V.34-199802-I}
\BIBentrySTDinterwordspacing

\bibitem{duan1997approaching}
L.~Duan, B.~Rimoldi, and R.~Urbanke, ``Approaching the {AWGN} channel capacity
  without active shaping,'' in \emph{Proc. IEEE Int. Symp. Inf. Theory (ISIT)},
  1997, p. 374.

\bibitem{ma2004coded}
X.~Ma and L.~Ping, ``Coded modulation using superimposed binary codes,''
  \emph{{IEEE} Trans. Inf. Theory}, vol.~50, no.~12, pp. 3331--3343, Dec. 2004.

\bibitem{cronie2010signal}
H.~S. Cronie, ``Signal shaping for bit-interleaved coded modulation on the
  {AWGN} channel,'' \emph{{IEEE} Trans. Commun.}, vol.~58, no.~12, pp.
  3428--3435, 2010.

\bibitem{legoff2005bit}
S.~Y. Le~Goff, B.~S. Sharif, and S.~A. Jimaa, ``Bit-interleaved turbo-coded
  modulation using shaping coding,'' \emph{{IEEE} Commun. Lett.}, vol.~9,
  no.~3, pp. 246--248, 2005.

\bibitem{legoff2007constellation}
S.~Y. Le~Goff, B.~K. Khoo, C.~C. Tsimenidis, and B.~S. Sharif, ``Constellation
  shaping for bandwidth-efficient turbo-coded modulation with iterative
  receiver,'' \emph{{IEEE} Trans. Wireless Commun.}, vol.~6, no.~6, pp.
  2223--2233, 2007.

\bibitem{khoo2006bit}
B.~K. Khoo, S.~Y. Le~Goff, B.~S. Sharif, and C.~C. Tsimenidis,
  ``Bit-interleaved coded modulation with iterative decoding using
  constellation shaping,'' \emph{{IEEE} Trans. Commun.}, vol.~54, no.~9, pp.
  1517--1520, 2006.

\bibitem{valenti2012constellation}
M.~C. Valenti and X.~Xiang, ``Constellation shaping for bit-interleaved {LDPC}
  coded {APSK},'' \emph{{IEEE} Trans. Commun.}, vol.~60, no.~10, pp.
  2960--2970, 2012.

\bibitem{bocherer2011operating}
G.~B\"ocherer and R.~Mathar, ``Operating {LDPC} codes with zero shaping gap,''
  in \emph{Proc. IEEE Inf. Theory Workshop (ITW)}, 2011.

\bibitem{bocherer2012capacity}
\BIBentryALTinterwordspacing
G.~B\"ocherer, ``Capacity-achieving probabilistic shaping for noisy and
  noiseless channels,'' Ph.D. dissertation, RWTH Aachen University, 2012.
  [Online]. Available:
  \url{http://www.georg-boecherer.de/capacityAchievingShaping.pdf}
\BIBentrySTDinterwordspacing

\bibitem{bocherer2011matching}
G.~B\"ocherer and R.~Mathar, ``Matching dyadic distributions to channels,'' in
  \emph{Proc. Data Compression Conf. (DCC)}, 2011, pp. 23--32.

\bibitem{bliss1981circuitry}
W.~Bliss, ``Circuitry for performing error correction calculations on baseband
  encoded data to eliminate error propagation,'' \emph{IBM Tech. Discl. Bull.},
  vol.~23, pp. 4633--4634, 1981.

\bibitem{blaum2007high}
M.~Blaum, R.~D. Cideciyan, E.~Eleftheriou, R.~Galbraith, K.~Lakovic,
  T.~Mittelholzer, T.~Oenning, and B.~Wilson, ``High-rate modulation codes for
  reverse concatenation,'' \emph{{IEEE} Trans. Magn.}, vol.~43, no.~2, pp.
  740--743, 2007.

\bibitem{kschischang1993optimal}
F.~R. Kschischang and S.~Pasupathy, ``Optimal nonuniform signaling for
  {G}aussian channels,'' \emph{{IEEE} Trans. Inf. Theory}, vol.~39, no.~3, pp.
  913--929, 1993.

\bibitem{cover2006elements}
T.~M. Cover and J.~A. Thomas, \emph{Elements of Information Theory},
  2nd~ed.\hskip 1em plus 0.5em minus 0.4em\relax John Wiley \& Sons, Inc.,
  2006.

\bibitem{ozarow1990capacity}
L.~Ozarow and A.~Wyner, ``On the capacity of the {G}aussian channel with a
  finite number of input levels,'' \emph{Information Theory, IEEE Transactions
  on}, vol.~36, no.~6, pp. 1426--1428, 1990.

\bibitem{blahut1972computation}
R.~Blahut, ``Computation of channel capacity and rate-distortion functions,''
  \emph{{IEEE} Trans. Inf. Theory}, vol.~18, no.~4, pp. 460--473, 1972.

\bibitem{arimoto1972algorithm}
S.~Arimoto, ``An algorithm for computing the capacity of arbitrary discrete
  memoryless channels,'' \emph{{IEEE} Trans. Inf. Theory}, vol.~18, no.~1, pp.
  14--20, 1972.

\bibitem{gallager2008principles}
R.~G. Gallager, \emph{Principles of Digital Communication}.\hskip 1em plus
  0.5em minus 0.4em\relax Cambridge University Press, 2008.

\bibitem{ungerboeck2002huffman}
G.~Ungerb\"ock, ``Huffman shaping,'' in \emph{Codes, Graphs, and Systems},
  R.~Blahut and R.~Koetter, Eds.\hskip 1em plus 0.5em minus 0.4em\relax
  Springer, 2002, ch.~17, pp. 299--313.

\bibitem{cai2007probabilistic}
N.~Cai, S.-W. Ho, and R.~Yeung, ``Probabilistic capacity and optimal coding for
  asynchronous channel,'' in \emph{Proc. IEEE Inf. Theory Workshop (ITW)},
  2007, pp. 54--59.

\bibitem{amjad2013fixed}
R.~A. Amjad and G.~B\"ocherer, ``Fixed-to-variable length distribution
  matching,'' in \emph{Proc. IEEE Int. Symp. Inf. Theory (ISIT)}, 2013.

\bibitem{baur2015arithmetic}
S.~Baur and G.~B\"ocherer, ``Arithmetic distribution matching,'' in \emph{Proc.
  Int. ITG Conf. Source Channel Coding (SCC)}, 2015.

\bibitem{schulte2015constant}
\BIBentryALTinterwordspacing
P.~Schulte and G.~B\"ocherer, ``Constant composition distribution matching,''
  \emph{arXiv preprint}, 2015. [Online]. Available:
  \url{http://arxiv.org/abs/1503.05133}
\BIBentrySTDinterwordspacing

\bibitem{wyner1975common}
A.~D. Wyner, ``The common information of two dependent random variables,''
  \emph{{IEEE} Trans. Inf. Theory}, vol.~21, no.~2, pp. 163--179, 1975.

\bibitem{imai1977new}
H.~Imai and S.~Hirakawa, ``A new multilevel coding method using
  error-correcting codes,'' \emph{{IEEE} Trans. Inf. Theory}, vol.~23, no.~3,
  pp. 371--377, May 1977.

\bibitem{gallager2013stochastic}
R.~G. Gallager, \emph{Stochastic processes: theory for applications}.\hskip 1em
  plus 0.5em minus 0.4em\relax Cambridge University Press, 2013.

\bibitem{ifabregas2010bit}
A.~G. {i F\`abregas} and A.~Martinez, ``Bit-interleaved coded modulation with
  shaping,'' in \emph{Proc. IEEE Inf. Theory Workshop (ITW)}, 2010, pp. 1--5.

\bibitem{kayhan2012constellation}
F.~Kayhan and G.~Montorsi, ``Constellation design for transmission over
  nonlinear satellite channels,'' in \emph{Proc. IEEE Global Telecommun. Conf.
  (GLOBECOM)}, 2012, pp. 3401--3406.

\bibitem{gray1953pulse}
F.~Gray, ``Pulse code communication,'' U. S. Patent 2\,632\,058, 1953.

\bibitem{li1997bit}
X.~Li and J.~A. Ritcey, ``Bit-interleaved coded modulation with iterative
  decoding,'' \emph{{IEEE} Commun. Lett.}, vol.~1, no.~6, pp. 169--171, 1997.

\bibitem{tanner_graph}
R.~M. Tanner, ``A recursive approach to low complexity codes,'' \emph{{IEEE}
  Trans. Inf. Theory}, vol.~27, no.~5, pp. 533--547, 1981.

\bibitem{richardson2001design}
T.~J. Richardson, M.~A. Shokrollahi, and R.~L. Urbanke, ``Design of
  capacity-approaching irregular low-density parity-check codes,'' \emph{{IEEE}
  Trans. Inf. Theory}, vol.~47, no.~2, pp. 619--637, 2001.

\bibitem{luby2001improved}
M.~G. Luby, M.~Mitzenmacher, M.~A. Shokrollahi, and D.~A. Spielman, ``Improved
  low-density parity-check codes using irregular graphs,'' \emph{{IEEE} Trans.
  Inf. Theory}, vol.~47, no.~2, pp. 585--598, 2001.

\bibitem{li2005bit}
Y.~Li and W.~E. Ryan, ``Bit-reliability mapping in {LDPC}-coded modulation
  systems,'' \emph{{IEEE} Commun. Lett.}, vol.~9, no.~1, pp. 1--3, 2005.

\bibitem{lei2008}
J.~Lei and W.~Gao, ``Matching graph connectivity of {LDPC} codes to high-order
  modulation by bit interleaving,'' in \emph{Proc. Allerton Conf. Commun.,
  Contr., Comput.}, Sept 2008, pp. 1059--1064.

\bibitem{gong2011improve}
L.~Gong, L.~Gui, B.~Liu, B.~Rong, Y.~Xu, Y.~Wu, and W.~Zhang, ``Improve the
  performance of {LDPC} coded {QAM} by selective bit mapping in terrestrial
  broadcasting system,'' \emph{{IEEE} Trans. Broadcast.}, vol.~57, no.~2, pp.
  263--269, 2011.

\bibitem{cheng2012exit}
T.~Cheng, K.~Peng, J.~Song, and K.~Yan, ``{EXIT}-aided bit mapping design for
  {LDPC} coded modulation with {APSK} constellations,'' \emph{{IEEE} Commun.
  Lett.}, vol.~16, no.~6, pp. 777--780, 2012.

\bibitem{hager2014optimized}
C.~H\"ager, A.~Graell~i Amat, A.~Alvarado, F.~Br\"annstr\"om, and E.~Agrell,
  ``Optimized bit mappings for spatially coupled {LDPC} codes over parallel
  binary erasure channels,'' in \emph{Proc. IEEE Int. Conf. Commun. (ICC)},
  Jun. 2014, pp. 2064--2069.

\bibitem{zhang_MET_journal}
L.~Zhang and F.~Kschischang, ``Multi-edge-type low-density parity-check codes
  for bandwidth-efficient modulation,'' \emph{{IEEE} Trans. Commun.}, vol.~61,
  no.~1, pp. 43--52, January 2013.

\bibitem{steiner_boecherer_2015}
\BIBentryALTinterwordspacing
F.~Steiner, G.~B\"ocherer, and G.~Liva, ``Protograph-based {LDPC} code design
  for bit-metric decoding,'' in \emph{IEEE Int. Symp. Inf. Theory (ISIT)},
  2015. [Online]. Available: \url{http://arxiv.org/abs/1501.05595}
\BIBentrySTDinterwordspacing

\bibitem{franceschini2006performance}
M.~Franceschini, G.~Ferrari, and R.~Raheli, ``Does the performance of {LDPC}
  codes depend on the channel?'' \emph{{IEEE} Trans. Commun.}, vol.~54, no.~12,
  pp. 2129--2132, 2006.

\end{thebibliography}
%\end{spacing}
%\clearpage
%\onecolumn
%\pgfplotsset{
%width=\columnwidth,
%height=0.5\textheight
%}
%\large
%\input{appendix}

\end{document}